\documentclass[lettersize,journal]{IEEEtran}
\usepackage{amsmath,amsfonts}
\usepackage{algorithmic}
\usepackage{algorithm}
\usepackage{array}
\usepackage[caption=false,font=normalsize,labelfont=sf,textfont=sf]{subfig}
\usepackage{textcomp}
\usepackage{stfloats}
\usepackage{url}
\usepackage{verbatim}
\usepackage{graphicx}
\usepackage{cite}
\usepackage{booktabs}
\usepackage{multirow}
\usepackage{float}
\begin{document}

\title{Domain Shift-oriented Machine Anomalous Sound Detection Model Based on Self-Supervised Learning }

\author{Jing-Ke Yan,Xin Wang,Qin Wang,Qin Qin,Huang-He LI,Peng-Fei Ye,Yue-Ping He,Jing Zeng ~\IEEEmembership{}
\thanks{This work is supported by General Project of Guangxi Natural Science Foundation (2019GXNSFAA245053), Guangxi Science and Technology Major Project (AA19254016), the National Natural Science Foundation of China (61862018), Guangxi Natural Science Foundation Project (2018GXNSFAA138084), Beihai city science and technology planning project (202082033), Beihai city science and technology planning project (202082023), Translation and Introduction of Guangxi Marine Culture under the Strategy of Maritime Power (2021KY0184),Guang xi graduate student innovation project(YCSW2021174).
	
Jing-Ke Yan is 1st author,Xin Wang and Qin Wang  and Qin Qin  are corresponding authors.

Xin Wang and Qin Wang  and Qin Qin are School of Marine Engineering,Guilin University of Electronic Technology.No. 1 Jinji Road, GuiLin, 541000, China. (email: 592499985@qq.com).
}

}

\markboth{Journal of \LaTeX\ Class Files,~Vol.~14, No.~8, August~2021}%
{Shell \MakeLowercase{\textit{et al.}}: A Sample Article Using IEEEtran.cls for IEEE Journals}


\maketitle

\begin{abstract}
Thanks to the development of deep learning, research on machine anomalous sound detection based on self-supervised learning has made remarkable achievements. However, there are differences in the acoustic characteristics of the test set and the training set under different operating conditions of the same machine (domain shifts). It is challenging for the existing detection methods to learn the domain shifts features stably with low computation overhead. To address these problems, we propose a domain shift-oriented machine anomalous sound detection model based on self-supervised learning (TranSelf-DyGCN) in this paper. Firstly, we design a time-frequency domain feature modeling network to capture global and local spatial and time-domain features, thus improving the stability of machine anomalous sound detection stability under domain shifts. Then, we adopt a Dynamic Graph Convolutional Network (DyGCN) to model the inter-dependence relationship between domain shifts features, enabling the model to perceive domain shifts features efficiently. Finally, we use a Domain Adaptive Network (DAN) to compensate for the performance decrease caused by domain shifts, making the model adapt to anomalous sound better in the self-supervised environment. The performance of the suggested model is validated on DCASE 2020 task 2 and DCASE 2022 task 2. 
\end{abstract}


\section{Introduction}
\IEEEPARstart{T}{he}  purpose of anomalous sound detection (ASD) study\cite{DBLP:conf/icassp/RusheN19} is to judge whether the sound made by a target machine is anomalous or not. In real situations, the anomalies are few but diverse, so it is challenging to collect exhaustive anomalous sounds but necessary to detect those anomalous sounds outside of the training set to monitor the machine's working condition. We need to use normal sound signals to train the model to make it learn a better probability distribution, fit anomalous sound signals, detect anomalous sound signals, and carry out fault detection of the machine. However, in real-world cases, normal sound signals would fluctuate due to the change of physical parameters of machine operating conditions, resulting in normal sounds being detected as anomalous sounds. Therefore, machine anomalous sound detection based on self-supervised learning has attracted extensive attention. Since 2020, DCASE Task 2 \cite{DBLP:journals/corr/abs-2106-04492}\cite{DBLP:conf/dcase/KoizumiKINNTPSE20} aims to develop a self-supervised machine anomalous sound detection model based on Deep Learning\cite{9772414}  to adapt to the new domain environment by training normal sounds of different domains. Recently, many Deep Learning based self-supervised anomalous sound detection methods have been applied to DCASE Task 2, mainly including Autoencoder based methods (AE methods) \cite{DBLP:conf/icassp/SuefusaNPTEK20}, and Flowed self-supervised density estimation method (Flow-based methods) \cite{DBLP:journals/corr/abs-2111-06539}, and self-supervised classification methods \cite{DBLP:conf/icassp/LiuGZW22}. AE-based methods first map the input features to low-dimensional vectors through the encoder and then restore the feature vectors in the low-dimensional vector space to the input features by the decoder. However, although those methods are effective in detection, their performances will significantly decrease if anomalous or unstable sound signals are involved in the training process. The Flow-based methods use the normalized flow to distinguish the normal and anomalous sound signals of the machine ID, thus improving the model's performance in detecting the anomalous sound of a single machine type. However, they must carry out the same distribution sampling for different machine IDs of each machine type, resulting in unstable detection efficiency. In order to better detect anomalous sound, numerous scholars have devoted themselves to the study of self-supervised anomalous machine sound detection. The self-supervised detection methods input all machine types and IDs into the model for training and use logarithmic likelihood or classification confidence to calculate the anomalous scores. Although their performance is much better than AE-based and Flow-based methods, their learning ability for the domain shifts feature of the sound signal is not strong, and their detection is unstable.

\begin{figure}[!t]
	\centering
	\includegraphics[width=2.5in]{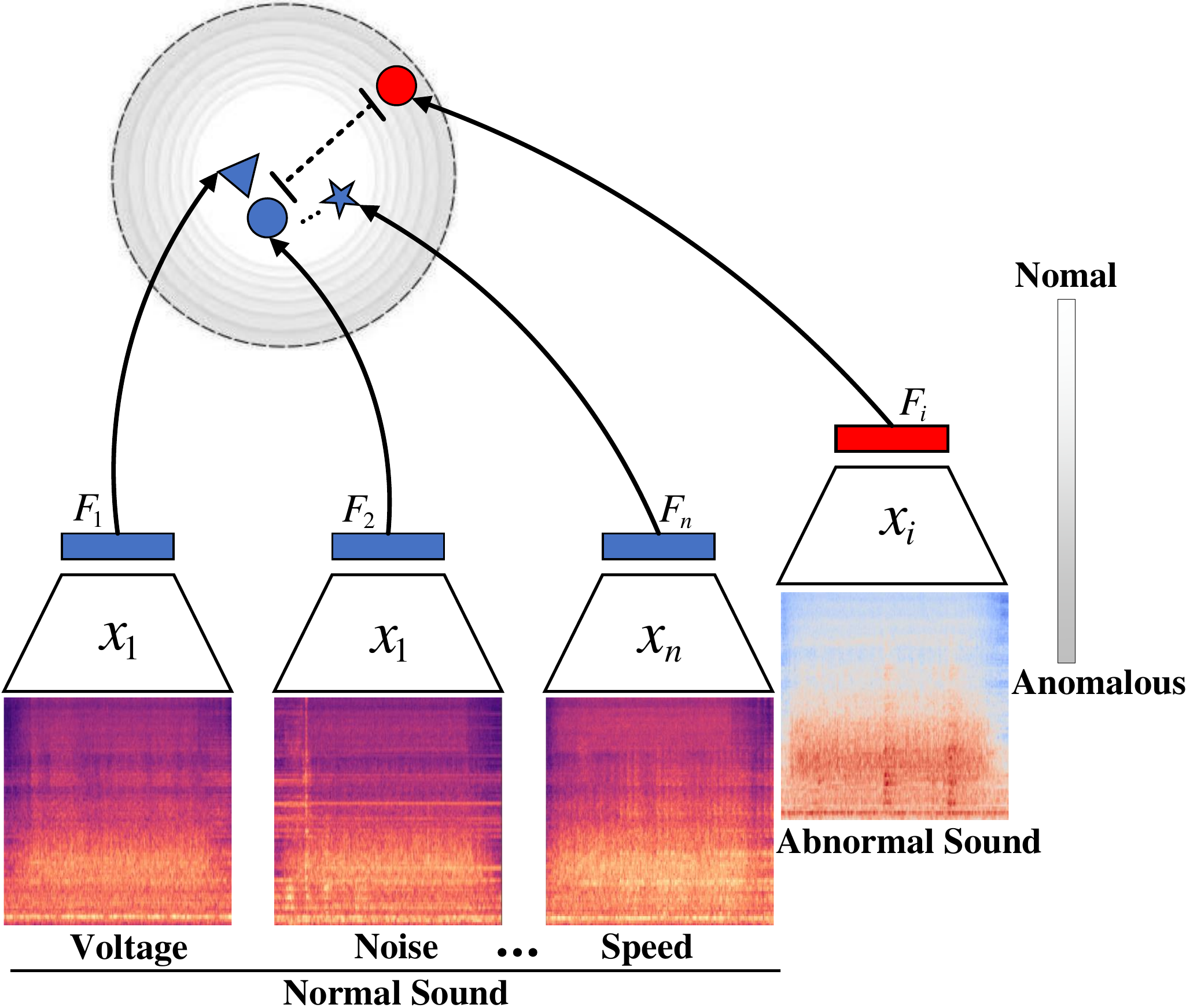}
	\caption{the model's mixed density estimation of the sound signal. Red indicates that the anomalous sound signal features distribute in the outer layer of the mixture density. Blue indicates that the normal sound signal features under different domain shifts (noise, rotational speed, voltage, etc.) distribute in the inner layer of the mixture density. (The dotted line indicates that the distance of mixture densities of the anomalous sound signal and the normal sound signal is as far as possible.)}
	\label{fig_1}
\end{figure}

To deal with these problems, we need to make the probability distribution of normal sound features of different domains shifts as close as possible, and the probability distribution of anomalous sound features as far as possible from that of normal sound features. Therefore, we propose a domain shift-oriented machine anomalous sound detection model based on self-supervised learning: TranSelf-DyGCN. In the model, the estimation of the probability distribution of the sound signal is equivalent to that of the mixture density of sound signal features, as shown in Fig.\ref{fig_1}. When we estimate mixture density, we aim to distribute the normal sound signal features under domain shifts in the innermost layer of the mixture density and the anomalous signal features in the outermost layer of the mixture density. Thus, we design a Feature Extractor Network, the dynamic graph convolutional networks (Dy-GCN), and Domain Adaptative Network (DAN) in our suggested model. Specifically, a time-domain Feature Extractor Network is designed to capture the time-domain features lost by the Mayer filter, helping the model extract the time-frequency domain features of sound signals. In addition, a new Transformer structure is advanced in the Feature Extractor Network to model the time-frequency domain features globally and locally, thus improving the stability of the Feature Extractor Network in the domain shifts environment. In the Dy-GCN, we design the Immutable Domain Transfer Graph Convolutional Network (Immutable Domain Transfer GCN) and the Variable Domain Transfer Graph Convolutional Network (Variable Domain Transfer GCN) to extract the relationship between the domain shifts features of the sound signal. Immutable Domain Transfer GCN mainly captures the coarse label dependencies between machine IDs on the dataset, thereby learning the relative semantic information of machine ID labels. The relative matrixes of the Variable Domain Transfer GCN are used to identify the semantic-aware features on each machine ID and to capture the subtle dependencies of domain shifts features in each machine ID. In addition, to further reduce the influence of domain shifts, we design a calculation method of second-order statistical difference in DAN to minimize the differences between the source domain features affected by domain shifts, and the target domain features not affected by domain shifts.

To sum up, the contributions of this paper are as follows:
\begin{enumerate}{}{}
\item{We propose a Feature Extractor Network to effectively compensate for the time-domain features lost when the model extracts time-frequency domain features . And it can  be used to model the long-distance dependence and local context information of time-frequency domain features;}

\item{We propose a GCN adaptively using  the relevant semantic features of sound signals to capture different domain shifts features in sound signals. Thus it is further enhanced the model's ability to distinguish domain shifts features;}

\item{We design a DAN which can minimize the variance of cross-domain learning features. And it can also enhance the models' stability in learning domain shifts features.}
\end{enumerate}
\section{Related Work}
In this section, we will discuss Autoencoder-based, Flow-based, and self-supervised classification methods, respectively, as shown in Fig.\ref{fig_10}.
\subsection{Autoencoder-based methods}
 Among Autoencoder-based(AE-based) methods, researchers \cite{DBLP:journals/taslp/WuCNY20}mainly focus on traditional AE-based methods and Variational Autoencoder (VAE)-based methods, as shown in Fig.\ref{fig_10} (a). Traditional AE-based methods \cite{DBLP:journals/pieee/RuffKVMSKDM21} utilize the model to reconstruct the input features and use the reconstructed features and input features to calculate the mean square error (MSE) as the judgment of the anomaly score. For example, Koizum et al. \cite{DBLP:journals/taslp/KoizumiSUKH19}  used Neyman-Pearson optimisation theory and autoencoder (AE) for unsupervised sound anomaly detection. The method first utilises the objective function of the Neyman-Pearson Lemma to statistically test the hypothesis of the sample. Huang et al. \cite{9632460} proposed an AE-based visual anomaly detection (VAD) method. The method first introduces an automatic encoder transformer (AT) to widen the anomaly score gap between normal and anomalous samples, and then uses the AE model to learn high-level semantic features of normal samples to obtain a potential representation of normal samples.
 
 The VAE-based methods \cite{DBLP:conf/icassp/BanerjeeG20} calculate the difference between the original input features and the reconstructed input features by using Approximate Posterior Probability Distribution and Kullback-Leibler Divergence of Prior Probability Distribution. For example, Yang et al. \cite{9662055} proposed a novel clustering method based on a spherical VAE. The method first deploys a hybrid model prior that can control the balance between the capacity of the decoder and the latent features. A dual VAE structure is then used to impose reconstruction constraints on the latent features and the corresponding noise. However, the VAE often collapses backwards when the decoder is used to parameterise an autoregressive model for sound sequence generation. Therefore Qian et al. \cite{9739027} designed InfoMaxHVAE to integrate the mutual aid information estimated by the neural network into a layered VAE, allowing the model to exhibit less post-collapse.
 
 The AE-based methods can detect stable sound signals well. However, suppose there are unstable sound signals in the training set. In that case, those methods are challenging to learn good feature probability distribution, and the reconstruction error is often huge, making the model difficult to fit. If anomalous sounds exist in the training set, the AE-based methods cannot be trained directly.
\subsection{Flow-based methods}
 Flow-based methods\cite{DBLP:journals/corr/abs-2111-06539} usually include a convolutional network-based density estimator and adaptive normalizer,as shown in Fig.\ref{fig_10} (b). The convolutional network-based density estimator handles the sound features under different domain shifts. At the same time, the adaptive normalizer can alleviate the differences between domains by scaling and shifting their feature distribution.  
 For example, Zhou et al. \cite{9735273}designed a novel inFlow prediction method that uses a normalised network to enhance the determinism of the model and thus extract multivariate correlation features between samples. In addition the method uses a density estimator to quantify the uncertainty of the prediction results, which can help to explain the behaviour of the model and the prediction results. However, inFlow-based methods are difficult to measure accurately from the latent variable space. liu et al. \cite{DBLP:conf/cvpr/LiuLGWL19} proposed a conditional adversarial generative flow (CAGlow), which learns mapping relations from the conditional space to the latent space in an adversarial manner. the CAGlow ensures the independence of different conditions. Although the use of adversarial networks for likelihood estimation in Flow models is a promising method for unsupervised anomaly detection, the method is susceptible to data smoothing. As a result, Dohi et al.  \cite{DBLP:conf/icassp/DohiEPTK2} assign higher weights to the model for target machine sound detection and lower weights for other machine sound detection from the same machine type.
 
 Flow-based methods learn the domain shifts features well, there may be problems in actual cases. They train different domains in each type separately. If the model is applied to a new machine ID for detection, the probability distribution of normal sounds usually changes due to the shifts of the target domain. The model cannot be used for detection directly. Hence, we need to optimize the model to make it adapt to this fluctuation. In addition, as the Flow-based methods need to train each machine type separately, they require much computational overhead, making them unavailable for some devices. 
 
\subsection{Self-supervised classification methods}
\begin{figure*}[ht]
	\centering
	\includegraphics[width=480px]{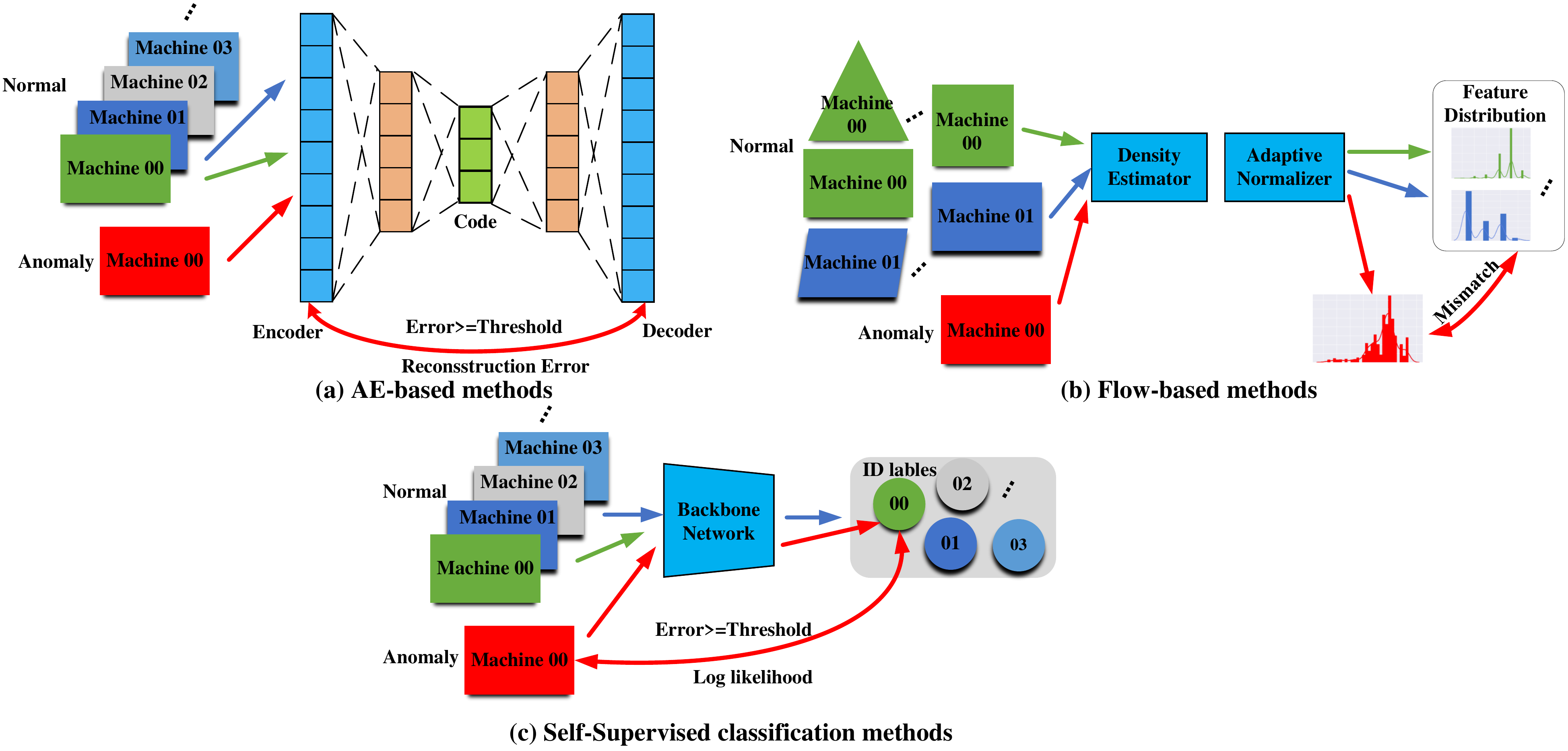}
	\caption{(a) AE-based methods (b) Flow-based methods  (c) Self-supervised classification methods.}
	\label{fig_10}
\end{figure*}

The self-supervised classification methods \cite{DBLP:conf/icassp/HojjatiA22} trains classifiers through metadata (i.e., machine type, machine IDs, as well as the labels under anomalous or normal condition) to predict normal machine IDs,,as shown in Fig.\ref{fig_10} (c). If the models based on self-supervised classification methods predict an anomalous sound, the classifier outputs the wrong machine ID. 
For example, Zhao et al. \cite{DBLP:conf/icassp/ZhaoZZMZ22} design a self-supervised pre-training method that uses the Swin Transformer  to learn meaningful features from a large number of easily collected unlabelled samples, and Liu et al. \cite{DBLP:conf/icassp/LiuGZW22} propose a self-supervised method based on spectral and temporal fusion for modelling normal machine sounds to improve the detection accuracy of abnormal sounds. Hojjati et al.  \cite{DBLP:conf/icassp/HojjatiA22} explored the application of contrast learning in anomaly detection. They used a contrast learning framework to learn the latent space in the sound and thus distinguish normal machine sounds from anomalous machine sounds.

Although the performance of these methods is better than that of AE and Flow-based methods, and the computational overhead is low, they are unstable when detecting some machines with domain shifts. Inspired by these methods, we suggest a brand-new method based on self-supervised classification, which can extract domain shifts features more stably.

\section{Methods}
This section species the newly-proposed model TranSelf-DyGCN. we first briefly introduce its architecture; then we elaborate its key components, including Feature Extractor Network, DyGCN, and DAN; finally, we introduce its Loss Function.
\subsection{A machine anomalous sound detection model: TranSelf-DyGCN}
\begin{figure*}[ht]
	\centering
	\includegraphics[width=\linewidth]{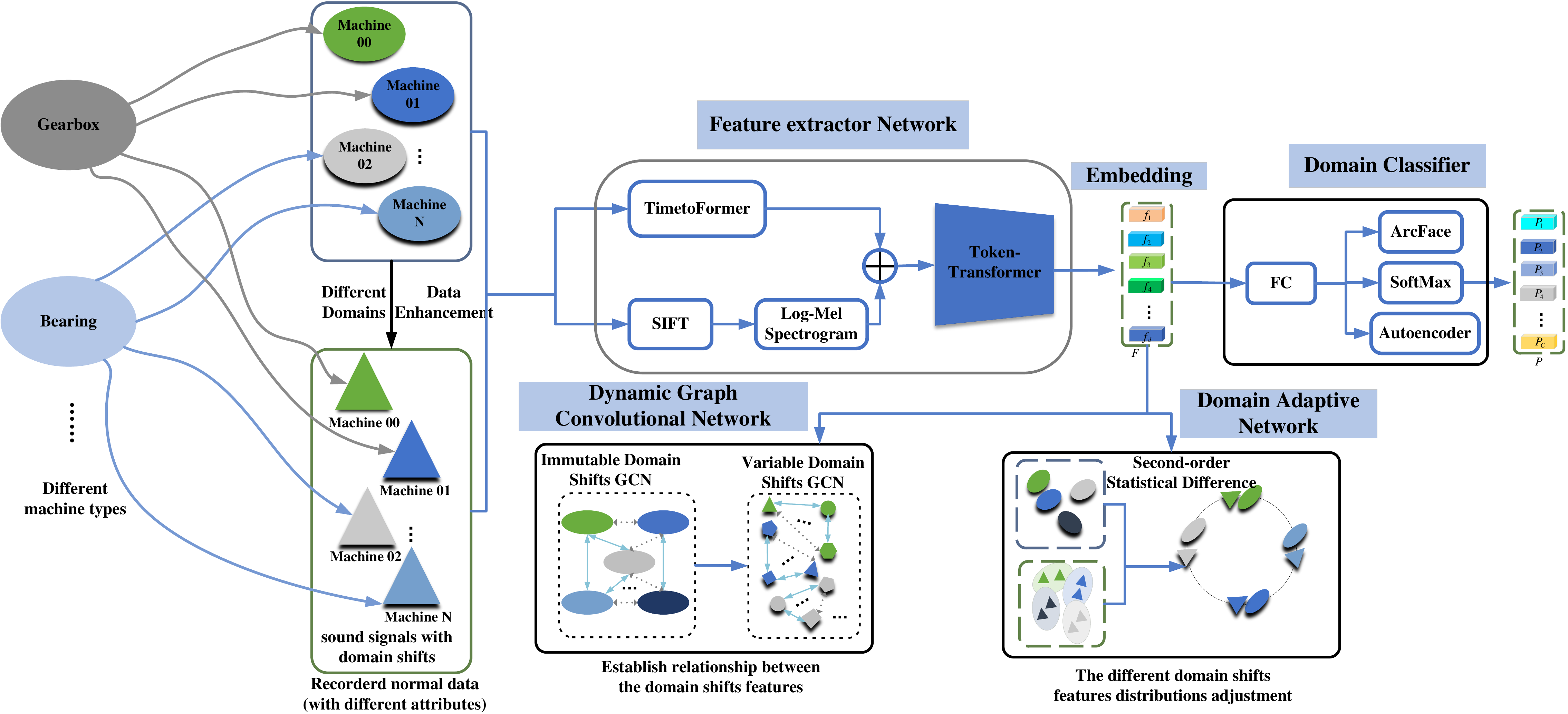}
	\caption{The overview of the TranSelf-DyGCN. The solid line in the Dynamic Graph Convolutional Network indicates closer relationships between categories or attributes. In contrast, the dotted line indicates the relationship between categories or attributes is distant. $ \oplus $ represents the addition of the time-frequency domain features extracted by STFT and Log-MEL Spectrogram and the anomaly lost time-domain features.}
	\label{fig_2}
\end{figure*}

In this paper, we need to solve the problem of self-supervised training without labeling anomalous sound signals and learn features that detect anomalous sounds in both source and target domains. To address those issues, we put forward the TranSelf-DyGCN model, as shown in Fig.\ref{fig_2}, to minimize the difference between the target and source domains. First, we select the sound signals with domain shifts of the same machine ID from the normal sound signals and transform them through the data enhancement to promote the model’s feature bias generation ability. Then, we input the normal sound signals and the transformed sound signal into the Feature Extractor Network in the TranSelf-DyGCN model (section \ref{FEN}) to get the feature vector $F$ of the sound signal. In Feature Extractor Network, Firstly, Short-time Fourier Transform (STFT) and Log-Mel Spectrogram are used to extract time-frequency domain features. Then the Timetoformer is used to capture the lost anomalous time-domain features caused by STFT and Log-MEL Spectrogram. The lost features are added and inputted into the Token-Transformer Network. At last, the Token-Transformer Network is used to model the local and global adaptive features’ spatial transformation and time-space relationship. The feature vector $F$ output by Feature Extractor Network is inputted into Dynamic Graph Convolutional Network (DyGCN, section \ref{DGCN}) and Domain Adaptive Network (DAN,section \ref{DAN}), and Domain Classifier, respectively. In DyGCN, we first use Immutable Domain Transfer GCN to improve the model's perception ability to detect coarse label dependencies between machine IDs. Then we use Variable Domain Transfer GCN to capture the subtle relationship between domain shifts features within machine IDs. In DAN, we calculate the second-order covariance differences between feature vectors $F$ under different domain shifts. In Domain Classifier (section \ref{LOSSFUNCTION}), we input the feature vector $F$into ArcFace \cite{DBLP:conf/cvpr/DengGXZ19}, Autoencoder (AE) \cite{DBLP:journals/tgrs/WangWZZ22}, and SoftMax, respectively. AE can train the model by using the errors of the reconstructed training samples and the input samples, making the model focus on the local features of the time-sequence signals.  ArcFace can improve the compactness of feature vector$F$ within machine IDs under domain shifts and the distinctiveness of feature vector$F$between machine IDs. The final output $P$ of the model can be calculated by SoftMax. It is worth noting that DyGCN, ArcFace, DAN, and AE are only used in training, and they update feature vectors $F$ through a gradient descent algorithm. DyGCN, DAN, ArcFace, and AE are not used when the model predicts. 

\subsection{ Feature Extractor Network \label{FEN}}

In the ASD tasks, to ensure the stability and consistency of anomalous sound signals under the same machine type, we input the time-frequency domain features of the sound into the Token-Transformer network and extract the feature vector   for fault detection.

\textbf{Short-time Fourier Transform (STFT) and Log-Mel Spectrogram:} To obtain the sound signal, we use the STFT and Log-Mel Spectrogram proposed by Liu \cite{DBLP:conf/icassp/LiuGZW22} et al. to transform the sound signal ${{x}_{v}}\in {{R}^{1\times T}}$and get the proper time-frequency domain features. Its expression is shown in Eq.(\ref{eq1}).
\begin{equation}
	\begin{array}{l}
	g(t,f) = {\int_{ - \infty }^{ + \infty } x _v}(\tau  - t) \cdot h(\tau  - t) \cdot {{\rm{e}}^{ - j2\pi f}}{\rm{d}}\tau \\
	{x_m} = \log \left( {{{\cal W}_i} \cdot {{\left\| {g(t,f)} \right\|}^2}} \right),
	\label{eq1}
   \end{array}
\end{equation}

where $h\left( t \right)$ is the window function, $g\left( t,f \right)$is the time-frequency domain features of time $t$and frequency $f$.${{\mathcal{W}}_{i}}\in {{R}^{L\times C}}$ represents the Mel filter group,$L$ stands for the dimension of spectrum features, and ${\rm{C}}$  represents the number of frequency boxes of the spectrum map obtained by STFT.${{x}_{m}}$ stands for the time-frequency domain features enhanced by Log-Mel Spectrogram.

\begin{figure*}[ht]
	\centering
	\includegraphics[width=350px]{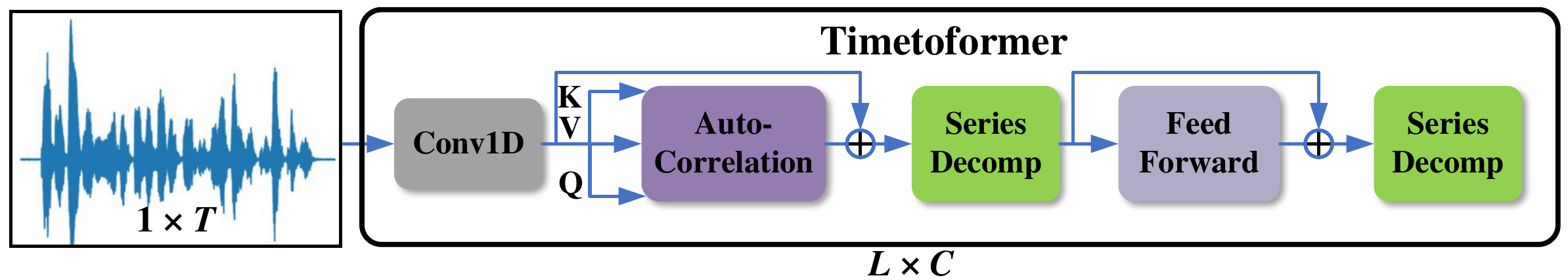}
	\caption{the structure of the Timetoformer network. $\oplus $represents the Residual connection.}
	\label{fig_3}
\end{figure*}

\textbf{Timetoformer:} the time-frequency domain features extracted by STFT and Log-Mel Spectrogram are susceptible to losing the anomalous time-sequence feature information. Liu et al. \cite{DBLP:conf/icassp/LiuGZW22} used the TgramNet network to extract time-sequence features from sound signals ${{x}_{v}}$ to compensate the lost time-sequence features in time-frequency domain features. Although TgramNet uses one-dimensional convolution with large cores to extract time-sequence features, it is difficult for one-dimensional convolution to find reliable interdependence of time-sequence features in long sequences. Recently, Qin et al. \cite{DBLP:journals/access/QinYWWLW21} used Transformer to capture the interdependence of time-sequence features. However, Transformer uses the sparse attention mechanism, which is quadratically complex in computation and cannot fully model the local and global relationship of time-sequence features of sound signals. Xu et al. \cite{DBLP:conf/nips/WuXWL21} thought that sub-sequences of time-sequence signals are usually similar at the same position in different cycles, which can be used to capture the time-sequence features. Inspired by this, we design the TimetoFormer network, as shown in Fig.\ref{fig_3}. This network uses the similarity of sub-sequences to capture anomalous time-sequence features with less computation. In the TimetoFormer network, 1-dimensional convolution is first used to encode the sound signal along the time dimension. Then, the Auto-Correlation Mechanism is used to aggregate similar sub-sequence information and replace the self-attention mechanism of the dot-product in the Transformer. This mechanism captures the internal time-sequence feature similarity of the encoded sound signal and reduces the computational complexity to $O(L\log L)$, whose formula is shown in Eq.(\ref{eq2}).
\begin{equation}
 	\begin{array}{l}
		{\tau _1}, \cdots ,{\tau _k} = \mathop {\arg  Topk }\limits_{\tau  \in \{ 1, \cdots ,L\} } \left( {{{\cal R}_{{\cal Q},{\cal K}}}(\tau )} \right)\\
		{\widehat {\cal R}_{{\cal Q},{\cal K}}}\left( {{\tau _i}} \right) =  SoftMax \left( {{{\cal R}_{{\cal Q},{\cal K}}}\left( {{\tau _i}} \right)} \right)\\
		\qquad where \qquad{\rm{ \mathit{i}}} \in 1,2, \cdots ,k\\
		Auto({\cal Q},{\cal K},{\cal V}) = \sum\limits_{i = 1}^k { Roll } \left( {{\cal V},{\tau _i}} \right){\widehat {\cal R}_{{\cal Q},{\cal K}}}\left( {{\tau _i}} \right)

	\end{array}
\label{eq2}
\end{equation}

where $arg Topk \left( \cdot  \right)$ represents the parameters of $k$ autocorrelation sub-sequences. $\mathcal{R}(\tau )$ represents the sub-sequence ${{\tau }_{1}},\cdots ,{{\tau }_{k}}$ obtained by the time window slides $k$ cycle lengths.  ${{\widehat{\mathcal{R}}}_{\mathcal{Q},\mathcal{K}}}$ is the autocorrelation weight between sequence $Q$ and $K$. $Auto(\mathcal{Q},\mathcal{K},\mathcal{V})$ means the sub-sequences of cycle alignment and phase similarity.

The encoded sound sequence ${{x}_{A}}\in {{R}^{L\times C}}$ of length dimension $L$ uses Series Decomp Block to extract the displayed high-level time-sequence features information, and the hidden high-level time-sequence features information in the middle of the model, as is shown in Eq.(\ref{eq3}).

\begin{equation}
	\begin{array}{l}
	{x_s} =  AvgPool ( Padding ({x_A}))\\
	{x_l} = {x_A} - {x_s}
	\end{array}
	\label{eq3}
\end{equation}

where ${{x}_{s}}\in {{R}^{L\times C}}$ represents the short-term fluctuation of time-sequence features, Padding represents a sliding window to pad 0 at both ends of the time-sequence features,AvgPool represents AvgPool Global average pooling, ${{x}_{l}}\in {{R}^{L\times C}}$ represents the periodic fluctuation of the time-sequence features. In addition, we add the Feed Forward Layer to the TimetoFormer network, enabling the network to cluster the extracted time-sequence features better. The Feed Forward Layer consists of a fully connected feedforward network.

\begin{figure*}[ht]
	\centering
	\includegraphics[width=\linewidth]{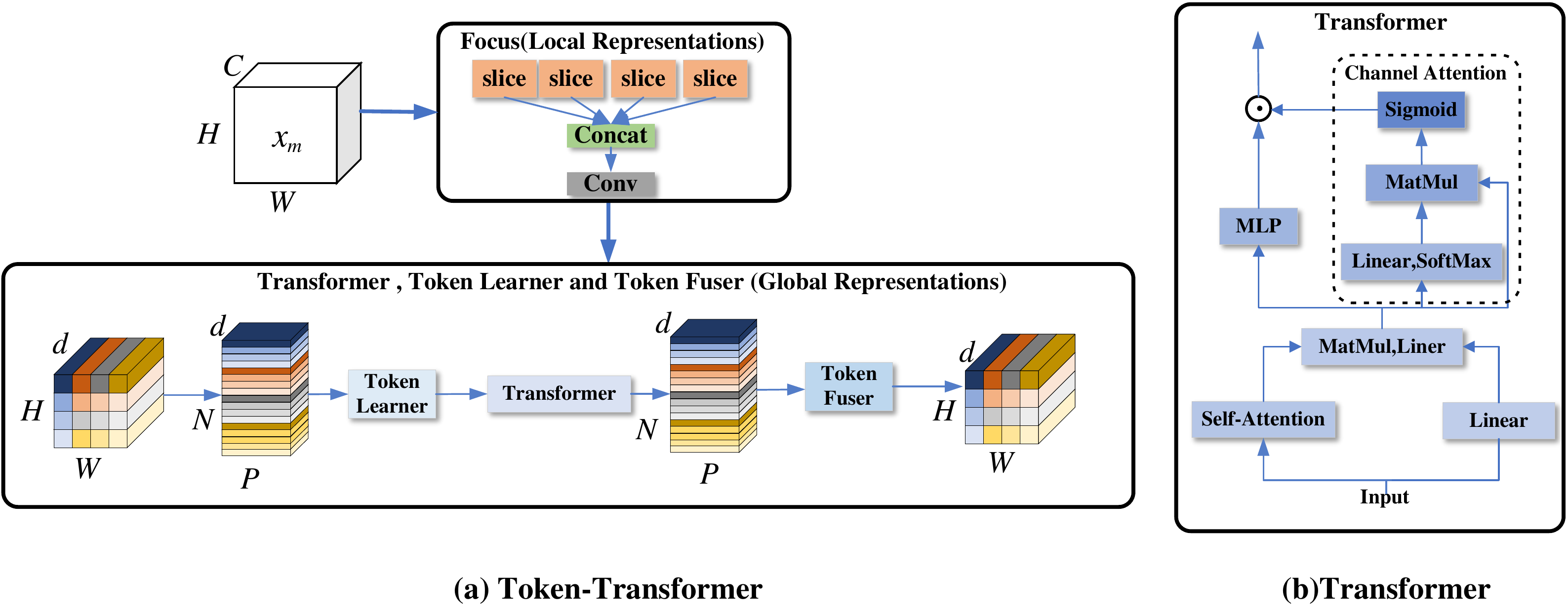}
	\caption{ (a) represents the architecture of the token-Transformer network (b) represents the structure of the Transformer, and $\odot $stands for the dot-product of matrix.}
	\label{fig_4}
\end{figure*}

\textbf{Token-Transformer:} When extracting time-frequency domain features, the Feature Extractor Network aims to fully model global and local feature information of input time-frequency domain features with fewer parameters. Therefore, we design the Token-Transformer network, as shown in Fig.\ref{fig_4}(a). In modeling local spatial information of time-frequency domain features, we adopt the Focus module, inspired by Ge et al. \cite{DBLP:journals/corr/abs-2107-08430}, to project the time-frequency domain features ${{x}_{m}}\in {{R}^{H\times W\times c}}$ to a high-dimensional space ${{x}_{L}}\in {{R}^{H\times W\times d}}\left( d>c \right)$. In order to enable token-Transformer to have spatial inductive global representation capability, we expand ${{x}_{L}}$ to ${{x}_{u}}\in {{R}^{P\times N\times d}}$, inspired by Mehta et al. \cite{DBLP:journals/corr/abs-2110-02178}, so that the model does not lose the sequence between patches of time-frequency features. where $P=wh$ and $N=HW/P$ is the numbers of patches, $h$ and $w$ are the height and width of the patch, respectively. In addition, in order not to lose the spatial order between tokens in patches, we use the Token Learner structure inspired by Ryoo et al. \cite{DBLP:journals/corr/abs-2106-11297}, which uses the spatial attention mechanism \cite{DBLP:conf/eccv/WooPLK18} to calculate the spatial weights between Tokens in Patches, and select $t$ important Tokens for modeling. Finally, we apply the Transformer to encode the relationship pair between Patches and the Tokens relationship pair within Patches, and model the global information relationship of time-frequency domain features, whose expression is shown in Eq.(\ref{eq4}).
\begin{equation}
	\begin{array}{l}
		x _ { G } ( p ) = \text { Transformer } \left( \text { TokenLearner } \left( x _ { U } ( p ) \right) \right)  \\ 
		\qquad where{\quad 1 \leq p \leq P}
	\end{array}
	\label{eq4}
\end{equation}

Additionally, inspired by Zhou et al. \cite{DBLP:journals/corr/abs-2204-12451}, we introduce a channel attention mechanism \cite{DBLP:journals/pami/HuSASW20} in the Multilayer Perceptron (MLP) module in Transformer, as shown in Fig.\ref{fig_4}(b), so as to facilitate the model to capture more domain shifts features. In this way, feature channels aggregated by Transformer can be considered more comprehensively, and more domain shifts features can be mined by using channel grouping information. We utilize the Token Fuser module \cite{DBLP:journals/corr/abs-2106-11297} to map ${{x}_{G}}={{R}^{P\times N\times t}}$ to ${{x}_{G}}={{R}^{P\times N\times d}}$ and fold ${{x}_{G}}$ to ${{x}_{F}}\in {{R}^{H\times W\times d}}$ as the feature map output of the Token-Transformer.
\subsection{ Dynamic Graph Convolutional Network  \label{DGCN}}
\begin{figure*}[ht]
	\centering
	\includegraphics[width=450px]{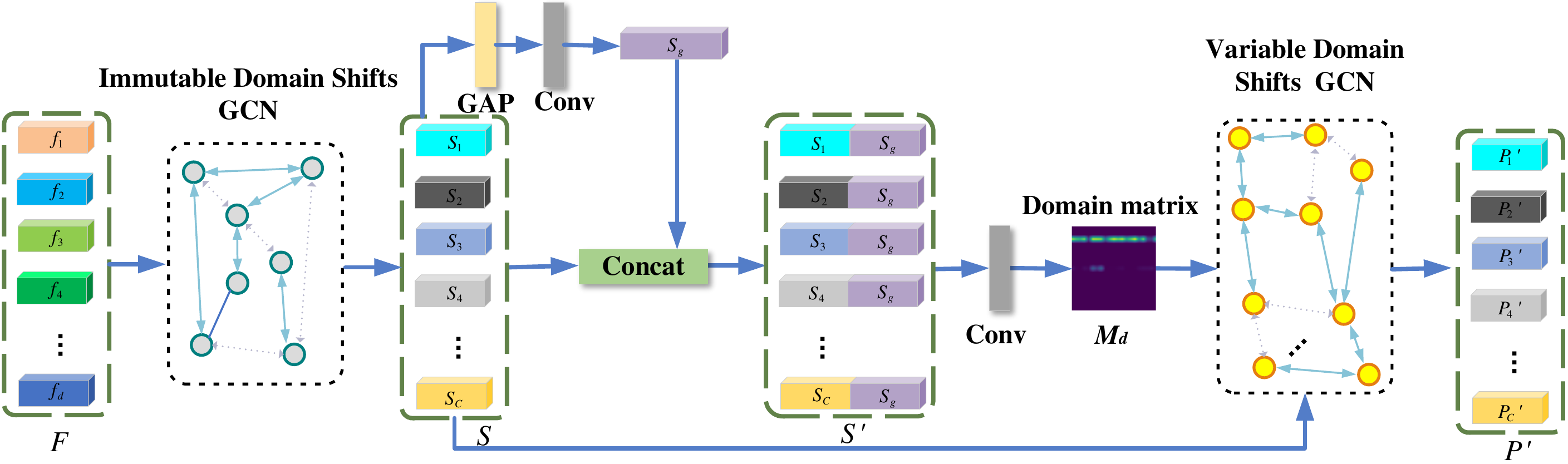}
	\caption{the overall architecture of the Dynamic Graph Convolutional Network. The solid line of Immutable Domain shifts GCN and Variable Domain shifts GCN indicates that the relationship between categories or attributes is close. The dotted line indicates that the relationship between categories or attributes is distant.}
	\label{fig_5}
\end{figure*}

In machine anomalous sound detection, how to effectively capture the dependencies between domain shifts features is an essential issue for self-supervised machine anomalous sound detection under domain shifts. Motivated by Ye et al. \cite{DBLP:conf/eccv/YeHPWQ20}, we advance a Dynamic Graph Convolutional Network to model the correlation between domains by considering the correlation of machine ID labels under different domain shifts, as shown in Fig.\ref{fig_5}. In this network, we design two modules: the Immutable Domain Transfer GCN (IDT-GCN) module and the Variable Domain Transfer GCN (VDT-GCN) module. We build a static GCN module in the IDT-GCN, which conducts content-aware modeling for the coarse relationship between machine IDs from the extracted feature map. Besides, we still build a dynamic GCN module in VDT-GCN, which captures the subtle dependencies of domain shifts features within machine IDs from the content-aware features output by the Immutable Domain Transfer GCN module.

\textbf{IDT-GCN:} In IDT-GCN, we use the traditional GCN to define the static GCN in our model and calculate the course-grained relationship of the machine ID labels. Given a set of features $F\in {{R}^{C\times d}}$ as input nodes, IDT-GCN uses relative matrix ${{M}_{s}}^{C\times C}$ and state update matrix ${{W}_{s}}^{d\times {{d}_{u}}}$ to calculate the coarse-grained relationship, whose calculation is shown in Eq.(\ref{eq5}).
\begin{equation}
	S = \delta ({M_s}F{W_s})
	\label{eq5}
\end{equation}

where ${{M}_{s}}$ and ${{W}_{s}}$ are randomly initialized by the gradient descent algorithm, ${{M}_{s}}$is a shared weight of all the time-frequency domain features of all sounds. Our purpose is to enable ${{M}_{s}}$ to capture the global coarse-grained classification dependency of different machine IDs. $\delta $ indicates the LeakyRelu activation function.

\textbf{VDT-GCN:} In the VDT-GCN module, we adaptively estimate the corresponding ${{M}_{d}}$ matrix based on the input feature $S$. Because different domain shifts feature of the ID of each machine type has different ${{M}_{d}}$ matrices, the ${{M}_{d}}$ matrix can enable the model to better extract features under different domain shifts. The calculation method of VDT-GCN is shown in Eq.(\ref{eq6}).

\begin{equation}
	\begin{array}{l}
		{P^\prime } = \delta \left( {{M_d}S{W_d}} \right),{\rm{ }}\\
		{\rm{where }}{M_d} =  Sigmoid \left( {{W_A}{S^\prime }} \right)
	\end{array}
	\label{eq6}
\end{equation}

where ${{W}_{d}}\in {{R}^{{{d}_{1}}\times {{d}_{2}}}}$ is the updated weight, and ${{W}_{A}}\in {{R}^{C\times {{d}_{1}}}}$ is the weight of the convolutional layer that constructs the dynamic correlation matrix ${{M}_{d}}$.$S'$ is obtained by connecting $S$ and ${{S}_{g}}\in {{R}^{d1}}$, and ${{S}_{g}}$ is calculated by the global average pooling layer (GAP) and the convolutional layer.

Finally, we use the cross-entropy Loss Function to update the feature vector $F$through the generated label $P'$, making the feature vector $F$ accurately capture the dependencies between the domain shifts features. The calculation is shown in Eq.(\ref{eq7}). It must be noted that Dy-GCN is only used to update features during training and is not used by the model during prediction.

\begin{equation}
	{\ell _{gcn}} = \sum\limits_{c = 1}^C {{y^c}} \log \left( {P'} \right) + \left( {1 - {y^c}} \right)\log \left( {1 - P'} \right)
	\label{eq7}
\end{equation}
\subsection{ Domain Adaptive Network  \label{DAN}}
In anomalous machine sound detection, there are different domain shifts in the training and test sets, resulting in their different mixture density distributions. The model predicts normal sound signals as anomalous sound signals. We hope the final learned features can also be applied well to the test set. Traditional solutions utilize relative entropy \cite{kapoor2018boosting}, Bregman divergence \cite{sprung2019upper}, and contrastive learning \cite{DBLP:conf/icml/RadfordKHRGASAM21} to estimate its distribution density. The above methods can learn the domain invariance of shallow features to a certain extent. However, they are still susceptible to domain shifts, complicated calculations, and low efficiency. Therefore, inspired by Sun et al. \cite{DBLP:conf/eccv/SunS16}, we design the Domain Adaptive Network to minimize the second-order statistical difference between source and target features.

We randomly extract sound signals from the same machine ID under different domain shifts from the training set. Then we perform data enhancement operations on the sound signals under different domain shifts, making the same machine ID have more different domain shifts features. We calculate the second-order covariance of feature vectors under different domain shifts. The calculated covariance uses the reverse gradient propagation algorithm to reduce the distance between feature vectors of different domain shifts. 

\textbf{Data Enhancement:} We utilize many transformations to generate more domain shifts sound signals. For example, Pitch Shift is used to increase or decrease the sound signals. Time Shifting is used to move sound signals forward or backward randomly. Time Stretch is used to change the speed of the sound signals. Gaussian white noise is injected into the sound signals to increase their robustness. FilterAugment \cite{DBLP:conf/icassp/NamKP22} applies different weights on the frequency band to simulate acoustic filters.

\textbf{Second-order covariance distance calculation:} When calculating the second-order statistic distance, we define the source domain feature as ${{F}_{S}}=\left\{ {{f}^{S}}_{i} \right\}$ and the target feature after domain shifts as ${{F}_{T}}=\left\{ {{f}^{T}}_{i} \right\}$, where $f\in {{R}^{d}}$. Therefore, we define the second-order covariance loss to calculate the distance between the source and target domains, as shown in Eq.(\ref{eq8}).

\begin{equation}
	\begin{array}{*{20}{l}}
		{{C_S} = \frac{1}{{{n_S} - 1}}\left( {F_S^ \top {F_S} - \frac{1}{{{n_S}}}{{\left( {{{\bf{1}}^ \top }{F_S}} \right)}^ \top }\left( {{{\bf{1}}^ \top }{F_S}} \right)} \right)}\\
		\begin{array}{l}
			{C_T} = \frac{1}{{{n_T} - 1}}\left( {F_T^ \top {F_T} - \frac{1}{{{n_T}}}{{\left( {{{\bf{1}}^ \top }{F_T}} \right)}^ \top }\left( {{{\bf{1}}^ \top }{F_T}} \right)} \right)\\
			{\ell _{seorco}} = \frac{1}{{4{d^2}}}\left\| {{C_S} - {C_T}} \right\|_F^2
		\end{array}
	\end{array}
	\label{eq8}
\end{equation}

where $C$ represents the covariance matrix of features, $n$ represents the number of input target and source domain features, and $d$ represents the dimension of the feature vector.

\subsection{Loss Function  \label{LOSSFUNCTION}}
Autoencoder Loss Function\cite{DBLP:journals/tgrs/WangWZZ22}(AE Loss) has been widely used and has achieved excellent performance in ASR tasks. In this paper, to enable the model to be adaptive to the sound features after domain shifts, we integrate the AE Loss into the model, as shown in Eq.(\ref{eq9}).

\begin{equation}
	{\ell _{AE}} = {\ell _{{\rm{MSE}}}}({\hat x_m} - {x_m})
	\label{eq9}
\end{equation}

where ${{x}_{m}}$ is the time-frequency domain feature extracted by STFT and Log-Mel Spectrogram, ${{\hat{x}}_{m}}$ is the Autoencoder reconstructed feature. ${{\ell }_{MSE}}$ represents the Mean-Squared Loss Function. Inspired by Jaehyeon et al.\cite{DBLP:conf/icml/KimKS21}, the reconstruction loss function in the model is defined through the time-frequency domain features extracted from SIFT and log-mel Spectrogram. Because the time-frequency domain features can significantly improve the perceptual quality of the model, more in line with the auditory perception system of the human ear. In addition, an audio waveform could be directly obtained from the projection mapping of the time-frequency domain features extracted from SIFT and log-mel Spectrogram, without any additional training parameters for the model.

In this paper, to make the sound features between domains more compact, we use the Additive Angular Margin Loss Function (AAM Loss) in Arcface to update features and obtain highly discriminative features. The calculation of AAM Loss is shown in Eq.(\ref{eq10}).

\begin{equation}
	{\ell _{arc}} =  - \frac{1}{N}\sum\limits_{i = 1}^N {\log } \frac{{{e^{s\left( {\cos \left( {{\theta _{{y_i}}} + m} \right)} \right)}}}}{{{e^{s\left( {\cos \left( {{\theta _{{y_i}}} + m} \right)} \right)}} + \sum\limits_{j = 1,j \ne {y_i}}^n {{e^{s\cos {\theta _j}}}} }}
	\label{eq10}
\end{equation}
AAM Loss distributes feature vector$F$ on a hyperplane of radius $s$to learn the similarity between domains within machine IDs and the difference between machine IDs and perform higher feature embedding optimization. $\theta $ is the angle between the weight $W$ and the feature. We replace the cross-entropy Loss Function with Poly Loss \cite{DBLP:journals/corr/abs-2204-12511} function. The Poly Loss Function is better than the cross-entropy Loss Function by adding the Taylor expansion to fit the probability distribution of sound signals. The calculation of the Poly Loss Function is shown in Eq.(\ref{eq11}).

\begin{equation}
	{\ell _{poly}} = {\ell _{CE}} + {\epsilon _1}\left( {1 - {P_C}} \right)
	\label{eq11}
\end{equation}

where ${{\ell }_{CE}}$ represents the cross-entropy Loss Function, ${{\epsilon }_{1}}$ represents the penalty coefficient term, and we set ${{\epsilon }_{1}}$ to 1 in this task. ${{P}_{C}}$ represents the real label. Finally, the Loss Function of our model is shown in Eq.(\ref{eq12}).

\begin{equation}
	{\ell _{total}} = {\ell _{AE}} + {\ell _{gcn}} + {\ell _{arc}} + {\ell _{seorco}} + {\ell _{poly}}
	\label{eq12}
\end{equation}
\section{Experiments and Analysis}
In this section, we introduce the experiment and validate our model. First, we introduce the detailed settings. Then, we compare the experimental results of our model and those of other models. Finally, we perform ablation experiments to elaborate the performance of the proposed model further.
\subsection{Experimental Settings}
\textbf{Experiment parameters:} The input sound signal of our model is about 10 seconds. If the sound clip is shorter than 10 seconds, we pad the input sound signal with 0. The length of the sampling window of the Log-Mel Spectrogram is set to 64ms, the frameshift is set to 32ms. The number of filters is set to 128, and the maximum frequency is set to 8000HZ, resulting in an input dimension of 128×311. We use the Admw optimizer \cite{DBLP:conf/iclr/LoshchilovH19} to train our model with the Weight decay set to 0.05. We adopt a cosine decay strategy. The maximum learning rate is set to 1.5e-4, and the minimum learning rate is set to 0. The margin of ArcFace is set to 0.7, and the scale is set to 30. The   of the Focal Loss Function is set to 0.25, and the   is set to 2. Batch Norm's eps is set to 1e-5, and Momentum is set to 0.1. The Negative Slope of the LeakyRelu activation function is set to 0.2. The model training graphics card is NVIDIA Tesla T4 16GB, and the batch size of the model is set to 30. Finally, we apply negative log probability to detect anomalous scores.

\textbf{Datasets:} Datasets from DCASE 2020 Task 2 and DCASE 2022 Task 2 are used to validate our method. There are six machine types on the DCASE 2020 Task 2 dataset. There are four different machine IDs in ToyConveyor. There are three different machine IDs in ToyCar, fan, pump, slider, and valve. There are seven different types of machines in DCASE 2022 Task 2, including ToyCar, ToyTrain, bearing, fan, gearbox, slider, and valve. Each machine type in DCASE 2022 Task 2 has three different machine IDs.

The main differences between the DCASE 2022 Task 2 datasets and DCASE2020 Task 2 datasets are as follows:
\begin{enumerate}{}{}
\item{ The training set and a test set of the DCASE 2022 Task 2 dataset have domain shifts under different conditions, while the training set and test set of the DCASE 2020 Task 2 dataset have domain shifts under the same conditions. }
\item{In the DCASE 2022 Task 2 test set, source domain data uninfluenced by the domain shifts and target domain data affected by domain shifts are mixed. The source domain data and the target domain data are not specified. However, the source and target domain data are specified in DCASE2020 Task 2. Therefore, it is more difficult to detect on DCASE2022 Task 2. }
\end{enumerate}

\textbf{Evaluation indicators:} We select Area Under Curve (AUC) and partial Area Under AUC (pAUC) to evaluate the model's performance. AUC is defined as the area enclosed by Receiver Operation Characteristics (ROC) curve and the coordinate axis. pAUC is AUC in the range of low false-positive rate (FPR) [0, P], and p is set to 0.1.
\subsection{Experimental results}
\textbf{DCASE 2020 Task 2:} To verify our model, we present the AUC and pAUC results of some advanced models tested on all machines on DCASE 2020 Task 2 in Table \ref{tab:table}. As can be seen from Table \ref{tab:table}, our proposed model improves AUC by 2.32\% and pAUC by 8.1\% compared with the current advanced STgram-MFN, demonstrating that our model has strong expressiveness of time-frequency domain features under different domain shifts. 

\textbf{DCASE 2022 Task 2:} we conduct experiments on DCASE 2022 Task \ref{tab:table2} and compare the results with some advanced models, whose experimental results are shown in Table \ref{tab:table2}. Table \ref{tab:table2} indicates that the average value of AUC and pAUC of our model on seven machine types reach 70.41\% and 62.46\%, respectively. The experimental results of TranSelf-DyGCN differ slightly from those of some advanced models on the Toycar, ToyTrain, and Fan machine type. For example, in the Toycar type, the AUC and pAUC are 7.89\% and 4.14\% lower than those of AE-based models. However, the overall detection performance of our model is better than AE-based models. The experimental results elaborate that our model can fully capture the time-frequency domain features of sound signals under domain shifts and can be well applied in anomalous sound detection tasks.

\begin{figure}[!t]
	\centering
	\includegraphics[width=3.5in]{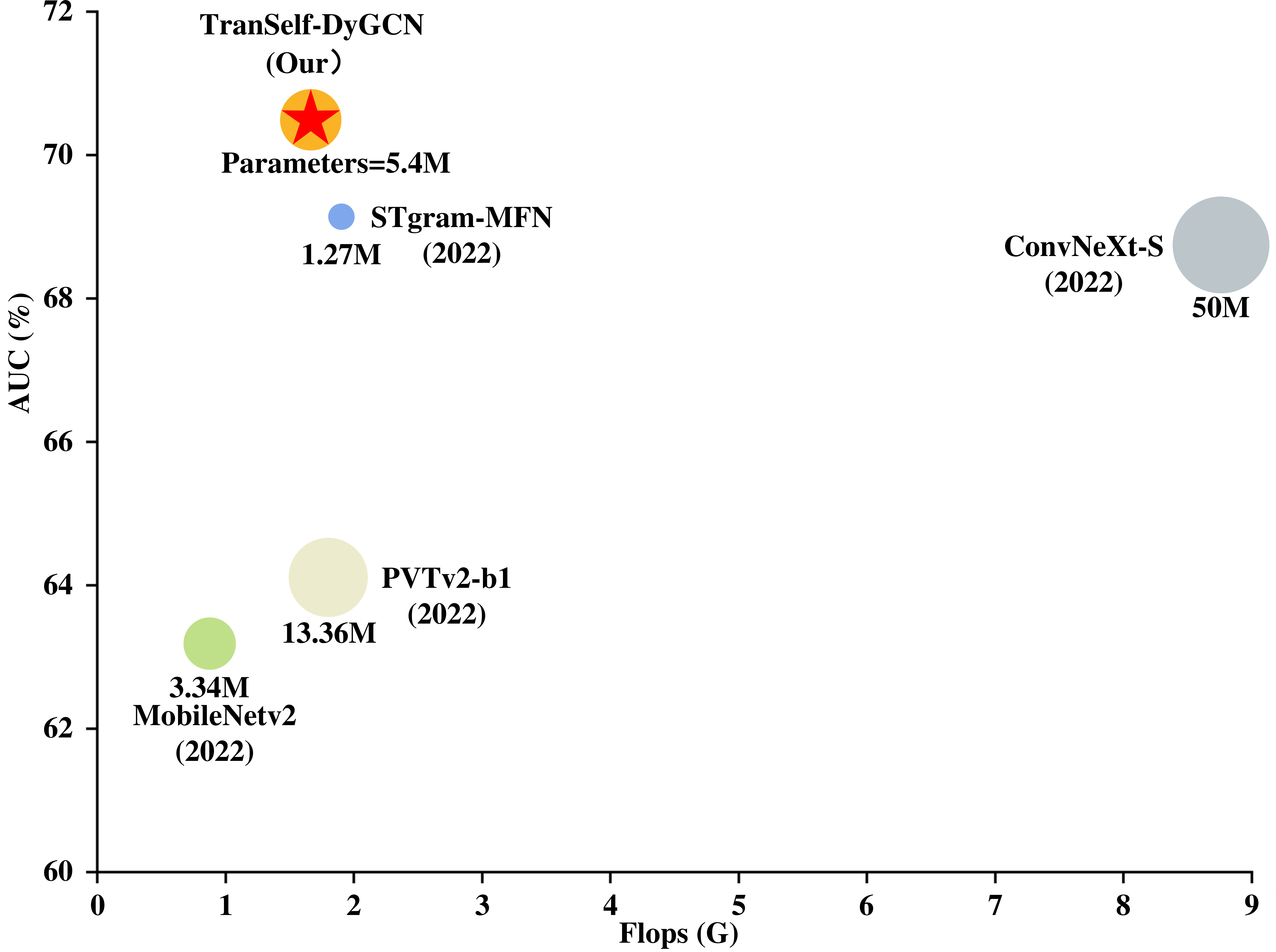}
	\caption{Comparison of  Flops, Parameters  and AUC on DCASE2022 Task2.}
	\label{fig_11}
\end{figure}

Finally, we visualize the Floating-point operations per second (Flops) and Parameters of the TranSelf-DyGCN model and other advanced models, as shown in Figure  \ref{fig_11}. As can be seen from Figure \ref{fig_11}, our model has Flops of 1.6G and Parameters of 5.4M. The experimental results not only show that our model can achieve better performance with a small computational overhead, but also that our model can run on smaller embedded devices.

\begin{table*}[]
	\caption{\label{tab:table}the comparison of AUC (\%) and pAUC (\%) of different models on different machine types on the DCASE2020 Task2 dataset. The best results are shown in bold.}
	\centering
	\setlength{\tabcolsep}{1mm}{
		\begin{tabular}{ccccccccccccccc}
		\hline
		\multirow{2}{*}{Models}     & \multicolumn{2}{c}{Fan}         & \multicolumn{2}{c}{Pump}        & \multicolumn{2}{c}{Slider}      & \multicolumn{2}{c}{Valve}       & \multicolumn{2}{c}{Toycar}      & \multicolumn{2}{c}{ToyConveyor} & \multicolumn{2}{c}{Average}     \\ 
		& AUC            & pAUC           & AUC            & pAUC           & AUC            & pAUC           & AUC            & pAUC           & AUC            & pAUC           & AUC            & pAUC           & AUC            & pAUC           \\ \hline
		IDNN{[}\cite{DBLP:conf/icassp/SuefusaNPTEK20}{]}(2020)          & 67.71          & 52.9           & 73.76          & 61.07          & 86.45          & 67.58          & 84.09          & 64.94          & 78.69          & 69.22          & 71.07          & 59.7           & 76.96          & 62.57          \\
		MobileNetv2{[}\cite{DBLP:conf/dcase/GiriTCHIK20}{]}(2020)   & 80.19          & 74.4           & 82.53          & 76.5           & 95.27          & 85.22          & 88.65          & 87.98          & 86.66          & 85.92          & 69.71          & 56.43          & 84.34          & 77.74          \\
		Glow\_Aff{[}\cite{DBLP:conf/icassp/DohiEPTK21}{]}(2021)     & 74.9           & 65.3           & 83.4           & 73.8           & 94.4           & 75             & 92.2           & 84.1           & 71.5           & 59             & 85.2           & 73.9           & 85.2           & 73.9           \\
		ESResNeXt{[}\cite{DBLP:conf/ijcnn/GuzhovRHD21}{]}(2021)     & 85.52          & 77.29          & 92.88          & 85.73          & 91.08          & 79.82          & 88.77          & 86.94          & 86.01          & 86.16          & 75.1           & 65.44          & 86.56          & 80.23          \\
		STgram-MFN{[}\cite{DBLP:conf/icassp/LiuGZW22}{]}(2022)     & 86.91          & 83.02          & 85.32          & 80.05          & 98.81          & 93.83          & 90.76          & 88.76          & 93.36          & 86.42          & 77.37          & 61.15          & 89.65          & 81.37          \\
		Our   Model & \textbf{88.09} & \textbf{88.06} & \textbf{87.31} & \textbf{85.42} & \textbf{99.29} & \textbf{96.93} & \textbf{91.97} & \textbf{89.47} & \textbf{94.32} & \textbf{90.17} & \textbf{82.87} & \textbf{76.48} & \textbf{91.97} & \textbf{89.47} \\ \hline
		\end{tabular} 
	}
\end{table*}

\begin{table*}[]
	\caption{\label{tab:table2}the comparison of AUC (\%) and pAUC (\%) of different models on different machine types on DCASE2022 Task 2 dataset. The best results are shown in bold.}
	\centering

	\setlength{\tabcolsep}{1mm}{
		\begin{tabular}{ccccccccccccccccc}
			\hline
		\multirow{2}{*}{Models}     & \multicolumn{2}{c}{ToyCar}         & \multicolumn{2}{c}{ToyTrain}        & \multicolumn{2}{c}{Fan}      & \multicolumn{2}{c}{Gearbox}       & \multicolumn{2}{c}{Bearing}      & \multicolumn{2}{c}{Slider} & \multicolumn{2}{c}{Valve}  & \multicolumn{2}{c}{Average}    \\
		& AUC            & pAUC           & AUC            & pAUC           & AUC            & pAUC           & AUC            & pAUC           & AUC            & pAUC           & AUC            & pAUC           & AUC            & pAUC        & AUC            & pAUC      \\ \hline
		AE{[}\cite{DBLP:journals/corr/abs-2205-13879}{]}(2021)          & \textbf{64.36} & 53.64 & 51.54 & 50.52 & 65.14 & 58.4  & 66.25 & \textbf{69.36} & 58.35 & 51.84 & 63.62 & 55    & 51.03 & 50.54 & 60.07 & 54.19 \\
		MobileNetv2{[}\cite{DBLP:conf/dcase/HaradaNTOYS21}{]}(2022)   & 57.34 & 53.71 & 55.98 & \textbf{52.78} & 51.9  & 58.08 & 62.06 & 54.77 & 59.95 & 54.57 & 86.24 & 71.42 & 66.67 & 63.56 & 63.12 & 57.35 \\
		
		ConvNeXt-S{[}\cite{DBLP:journals/corr/abs-2201-03545}{]}(2022)    & 59.51 & 50.12 & 55.97 & 51.57 & 68.03 & \textbf{63.17} & 71.34 & 56.29 & 67.09 & 63.21 & 78.64 & 60.89 & 80.09 & 69.34 & 68.84 & 59.01 \\
		 
		PVTv2-b1{[}\cite{DBLP:journals/cvm/WangXLFSLLLS22}{]}(2022)    & 59.36 & 49.28 & 54.69 & 51.19 & 64.17 & 59.71 & 64.46 & 54.19 & 59.9  & 57.61 & 82.64 & 71.8  & 62.78 & 63.17 & 64.04 & 58.13 \\
		 
		EdgeVITs{[}\cite{DBLP:journals/corr/abs-2205-03436}{]}(2022)    & 54.89 & \textbf{54.35} & 52.4  & 51.05 & 60.36 & 52.54 & 65.05 & 54.31 & 56.21 & 52.61 & 70.12 & 58.7  & 51.25 & 49.42 & 58.61 & 53.28 \\
		STgram-MFN(2022)     & 53.48 & 50.56 & 55.5  & 49.7  & 67.05 & 58.12 & 71.57 & 62.94 & 75.79 & 64.14 & 85.09 & 69.21 & 74.48 & 68.6  & 69.05 & 60.47 \\ 
		Our   Model & 56.47 & 49.5  & \textbf{56.5} & 50.08 & \textbf{70.21} & 62.23 & \textbf{75.35} & 65.66 & \textbf{67.19} & \textbf{69.01} & \textbf{86.74} & \textbf{78.87} & \textbf{80.15} & \textbf{71.68} & \textbf{70.41} & \textbf{62.46}  \\ \hline
		\end{tabular}
	}

\end{table*}
 \subsection{Ablation experiment}
In this part, we perform ablation experiments on the DCASE2022 Task 2 dataset to validate each key component of TranSelf-DyGCN, including Loss Function, Timetoformer, Token-Transformer, Dynamic Graph Convolutional Network (DyGCN), Domain Adaptive Network (DDN).

\textbf{Loss Function:} We first verify the effect of the Loss Function on our model’s prediction results. We compare the Autoencoder Loss Function (AE Loss), the Additive Angular Margin Loss Function (AAM Loss), the cross-entropy Loss Function, the Focal Loss Function, and the Ploy Loss Function. The results are shown in Table \ref{tab:table3}, which indicates that when AAM, AE, and Poly Loss Functions are involved in the model, the AUC and pAUC reach 70.41\% and 62.46\%, respectively. The results are superior to those of other Loss Functions, signifying that AAM, AE, and Ploy Loss Functions can make the model adapt to domain shifts better.

\begin{figure}[!t]
	\centering
	\includegraphics[width=2.5in]{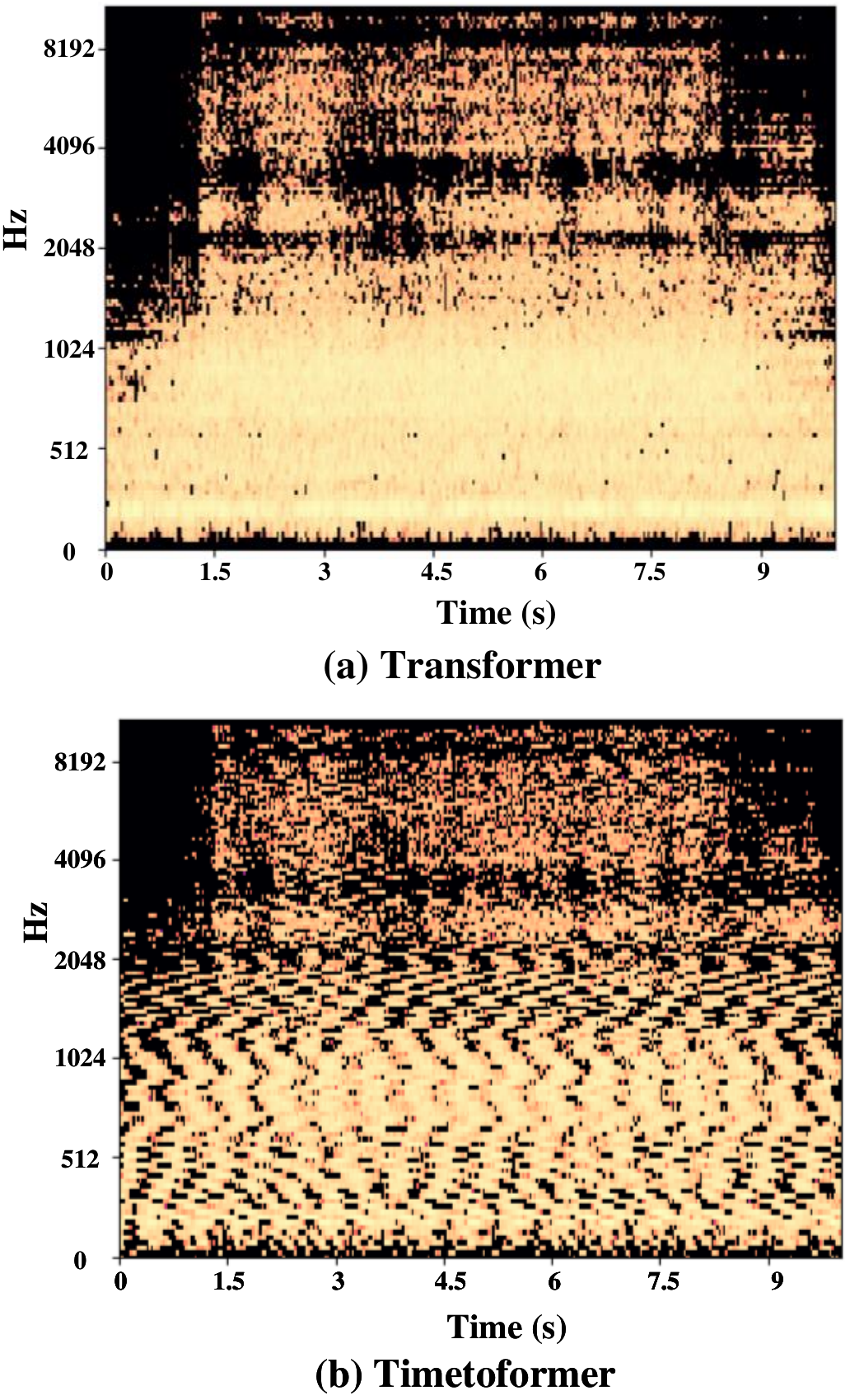}
	\caption{ shows the visualization of the spectrograms of Transformer and Timetoformer. (a) is the spectrogram of the Transformer. (b) is the spectrogram of Timetoformer.}
	\label{fig_6}
\end{figure}

\textbf{Time-frequency domain Feature Extraction Network:}
 To explore the role of the Timetoformer, we compare its effect with other time-sequence feature extraction models in Table \ref{tab:table4}. These results show that when Timetoformer is involved, AUC and pAUC increase by 1.45\% and 1.5\%, respectively, demonstrating that Timetoformer can capture anomalously lost time-sequence features well. Finally, we visualize the spectrogram of the sound signals extracted by models with Timetoformer and Transformer, respectively. As shown in Fig. \ref{fig_6}, we find out that the time-frequency domain features extracted by the model with Timetoformer are more periodic and have richer semantic information than those of the model with Transformer.

\begin{figure*}[ht]
 	\centering
 	\includegraphics[width=400px]{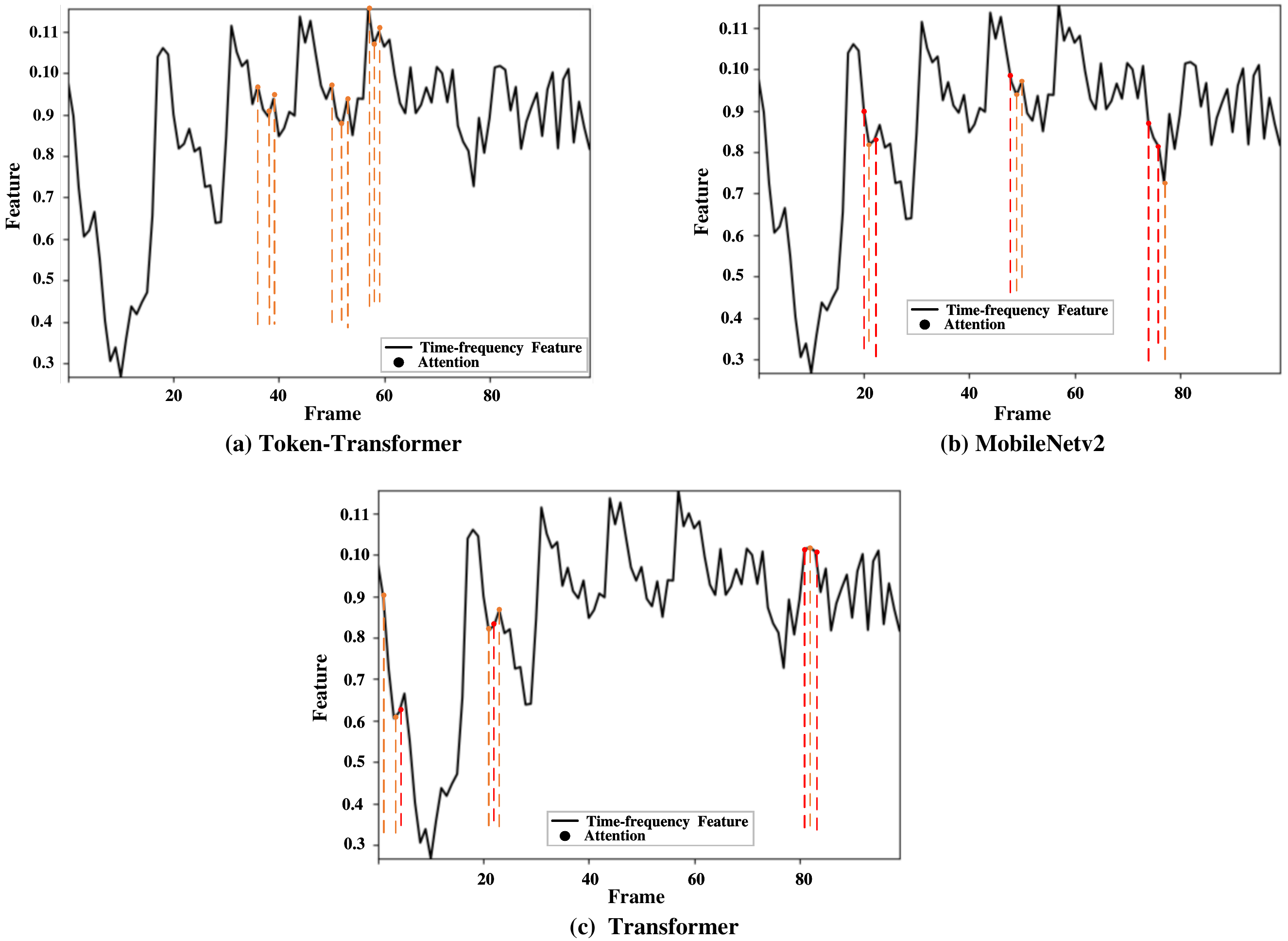}
 	\caption{the visualization of the time-frequency domain features of each model. The Orange and Red lines represent the attention weights for the current frame. The Orange line indicates the correct attention weight for the current frame, and the Red line indicates the wrong attention weight for the current frame.}
 	\label{fig_7}
 \end{figure*}

 In addition,To explore the Token-Transformer network’s ability to model time-frequency domain features, we compare different time-frequency domain Feature Extraction networks in Table \ref{tab:table5}. The third row of Table \ref{tab:table5} shows that the AUC and the pAUC of Token-Transformer increase by 3.4\% and 1.24\% compared with MobileNetv2, respectively. The fourth row of Table \ref{tab:table5} signifies that when the Token Learner(TL) is added to Token-Transformer, its AUC increases by 1.03\%, and the pAUC increases by 0.55\%. These results indicate that Token Learner can fully encode the spatial sequence of Patch sequence pairs in time-frequency domain features and the spatial sequence of Token sequence pairs in Patch. The fifth row of Table \ref{tab:table5} shows that when the channel attention mechanism(SE) is added to Toke-Transformer, its AUC increases by 0.61\%, and pAUC increases by 0.6\%. This increase demonstrates that integrating the channel attention mechanism into Token-Transformer can make it dig out more domain shifts features. Finally, the Token-transformer with channel attention mechanism and Token Learner, MobileNetv2, and Transformer model the time-frequency domain feature information dependencies, and the results are visualized, as is shown in Fig.\ref{fig_7}. The results elaborate that compared with MobileNetv2 and Transformer, Token-Transformer can capture the rising and falling process of time-frequency domain features of each cycle more accurately, with few missed and false detection. All of these indicate that Token-Transformer can better capture global and local detail features. 

\begin{figure}[!t]
 	\centering
 	\includegraphics[width=2.5in]{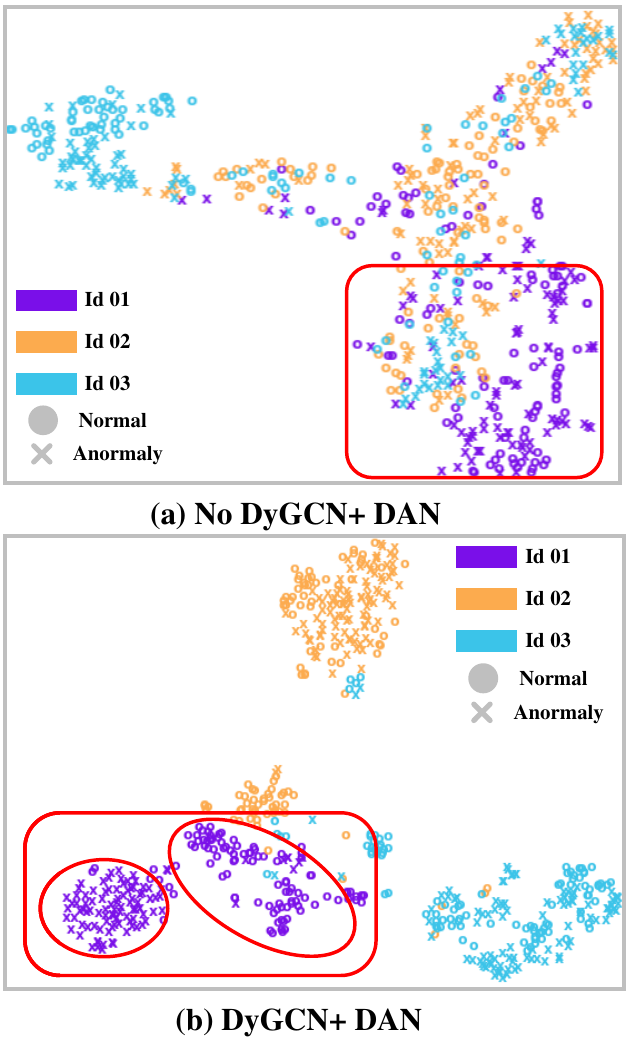}
 	\caption{t-SNE is used to visualize the feature maps extracted by Tranself-DyGCN. Different colors represent different machine IDs.}
 	\label{fig_8}
 \end{figure}

\textbf{DyGCN and DAN:} To investigate the role of DyGCN and DAN, we conduct experiments on TranSelf-DyGCN with and without DyGCN and DAN, whose results are shown in Table \ref{tab:table6}. As can be seen from the Table \ref{tab:table6}, The AUC and pAUC of the model with DyGCN increase by 1.64\% and 1.98\% compared with the model without DyGCN, which illustrate that the model with DyGCN can use the content-aware features between machine IDs to capture the subtle dependencies between domain shifts features. As can be seen from the fourth line of Table 6, the AUC and pAUC of the model with DAN increase by 2.36\% and 0.99\% compared with that of the model without DAN, which means that DAN can reduce the distance between the domain shifts features. At last, we compare the difference of features extracted from sound signals by the model with DyGCN and DAN and the one without DyGCN and DAN and use T-distributed stochastic neighbor embedding (T-SNE) to cluster and visualize the results, as is shown in Fig.\ref{fig_8}. Figure \ref{fig_8}(a) shows that the features extracted by the model without DyGCN and DAN overlap. In contrast, the features extracted by the model with DyGCN and DAN in Figure \ref{fig_8}(b) are easier to distinguish, indicating that the model with DyGCN and DAN can fully extract the domain shifts features.

\begin{table}[!ht]
	\caption{\label{tab:table3}the comparison of AUC (\%) and pAUC (\%) of models with different Loss Functions and the best results are shown in bold.}
	\centering
	\begin{tabular}{ccc}
		\hline
		Loss Function & AUC & pAUC  \\ \hline
		Cross Entropy Loss & 65.97 & 60.08  \\ 
		Focal Loss\cite{DBLP:conf/iccv/LinGGHD17} & 65.81 & 58.97  \\ 
		Poly Loss\cite{DBLP:journals/corr/abs-2204-12511} & 66.49 & 60.75  \\ 
		AAM Loss & 68.93 & 61.38  \\
		AAM+AE Loss & 69.99 & 62.01  \\
		AAM+AE+Poly Loss & \textbf{70.41} & \textbf{62.46}  \\ \hline
	\end{tabular}
\end{table}
\begin{table}[!ht]
	\caption{\label{tab:table4}the comparison of AUC (\%) and pAUC(\%) of models with different time-sequence Feature Extractor Networks and the best results shown in bold. ("-" means not to use any time-sequence Feature Extractor Network)}
	\centering
	\begin{tabular}{ccc}
		\hline
		Network & AUC & pAUC  \\ \hline
		- & 68.82 & 61.05  \\ 
		TgramNet\cite{DBLP:conf/icassp/LiuGZW22}  & 68.88 & 61.10  \\ 
		Transformer\cite{DBLP:conf/nips/VaswaniSPUJGKP17}  & 68.96 & 60.96  \\ 
		Timetoformer & \textbf{70.41} & \textbf{62.46}  \\ \hline
	\end{tabular}
\end{table}
 
\begin{table}[!ht]
	\caption{\label{tab:table5}the comparison of AUC (\%) and pAUC (\%) of models with different time-frequency domain Feature Extractor Networks and the best results are shown in bold. }
	\centering
	\begin{tabular}{ccc}
		\hline
		Network & AUC & pAUC  \\ \hline
		Transformer\cite{DBLP:conf/nips/VaswaniSPUJGKP17} & 65.09 & 57.48  \\ 
		MobileNetv2\cite{DBLP:conf/cvpr/SandlerHZZC18} & 65.37 & 60.07  \\ 
		Token-Transformer & 68.77 & 61.31  \\ 
		Token-Transformer+TL & 69.8 & 61.86  \\ 
		Token-Transformer+TL+SE & \textbf{70.41} & \textbf{62.46}  \\ \hline
	\end{tabular}
\end{table}
\begin{table}[!ht]
	\caption{\label{tab:table6}the AUC (\%) and pAUC (\%) of models with and without DyGCN and DAN, and the best results are shown in bold. (where "-" means that DyGCN and DAN are not added to the model) }
	\centering
	\begin{tabular}{ccc}
		\hline
		Network & AUC & pAUC  \\ \hline
		- & 66.41 & 59.49  \\ 
		DyGCN & 68.05 & 61.47  \\
		DAN+DyGCN & \textbf{70.41} & \textbf{62.46}  \\ \hline
	\end{tabular}
\end{table}

\section{Conclusion}
This paper suggests a domain shift-oriented machine anomalous sound detection model (TranSelf-DyGCN) based on self-supervised learning, which can detect machines' anomalous sound well in different domains. Its optimizations include the following aspects: Firstly, we advance a network that can fully model time-frequency domain features, which can capture the time-domain features lost by the Log-Mel spectrum, and can thoroughly learn the local and global dependencies of time-frequency domain features by using standard convolution and Transformer. This design dramatically improves the model's ability to extract domain shifts features. Then, we propose a Dynamic Graph Convolutional Network, which can fully capture the content-aware feature between machine IDs and use them to capture the relationship of domain shifts feature within machine IDs. It deepens the model's sensitivity to the domain transfer features and enhances its sensitivity to domain shifts features. Finally, we raise a Domain Adaptive Network, which uses second-order statistics to calculate the distance between domain shifts features to minimize the distance between domain shifts features and improve the model's stability of extracting domain shifts features. We conduct experiments on DCASE Task 2 to validate the excellent performance of our model. 

Our work still needs further improvement. For example, the following aspects are not discussed exhaustively. Whether the model proposed in this paper is suitable for other unsupervised detection tasks. Whether it can be reparametrized to speed up its reasoning time, and whether we can fully use transfer learning and train it with a small amount of data to achieve a good anomaly detection effect. In the future, we will continue to apply unsupervised learning based on TranSelf-DyGCN to other tasks (such as natural language and images, etc.) to increase the model's scalability.

\bibliographystyle{IEEEtran}
\bibliography{sample.bib}

\begin{IEEEbiography}
	[{\includegraphics[width=1in,height=1.25in,clip,keepaspectratio]{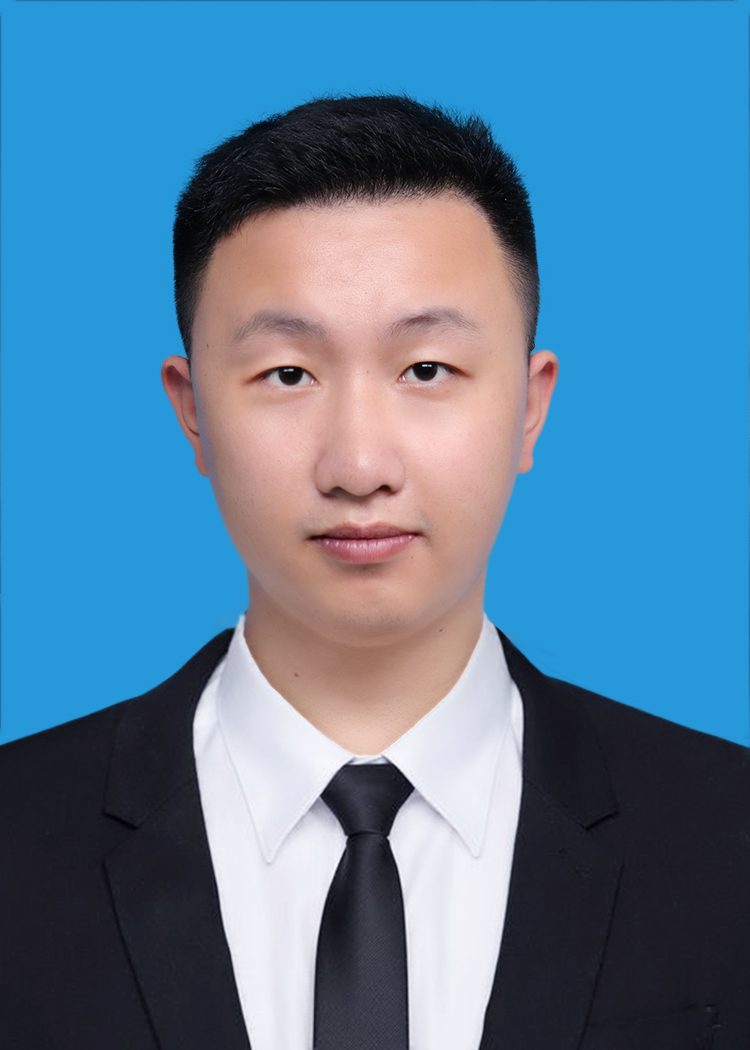}}]{JING-KE YAN}  (e-mail:592499985@qq.com) is currently a postgraduate student in the school of Marine Engineering of Guilin University of Electronic Technology. His research interests include image processing, deep learning, knowledge graph, reinforcement learning, natural language.
\end{IEEEbiography}
\vspace{-15 mm}
\begin{IEEEbiography}[{\includegraphics[width=1in,height=1.25in,clip,keepaspectratio]{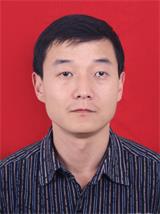}}]{XIN WANG}  (e-mail: 928311457@qq.com)is an associate professor, and master’s supervisor. His main research directions: image processing, network information security, Internet of Things, data mining. He is the main participant of National Natural Science Foundation of China, and the principal investigator of General Project of Guangxi Natural Science Foundation, Guangxi Science and Technology Major Project, Guangxi government planning project and etc.
\end{IEEEbiography}
\vspace{-15 mm}
\begin{IEEEbiography}[{\includegraphics[width=1in,height=1.25in,clip,keepaspectratio]{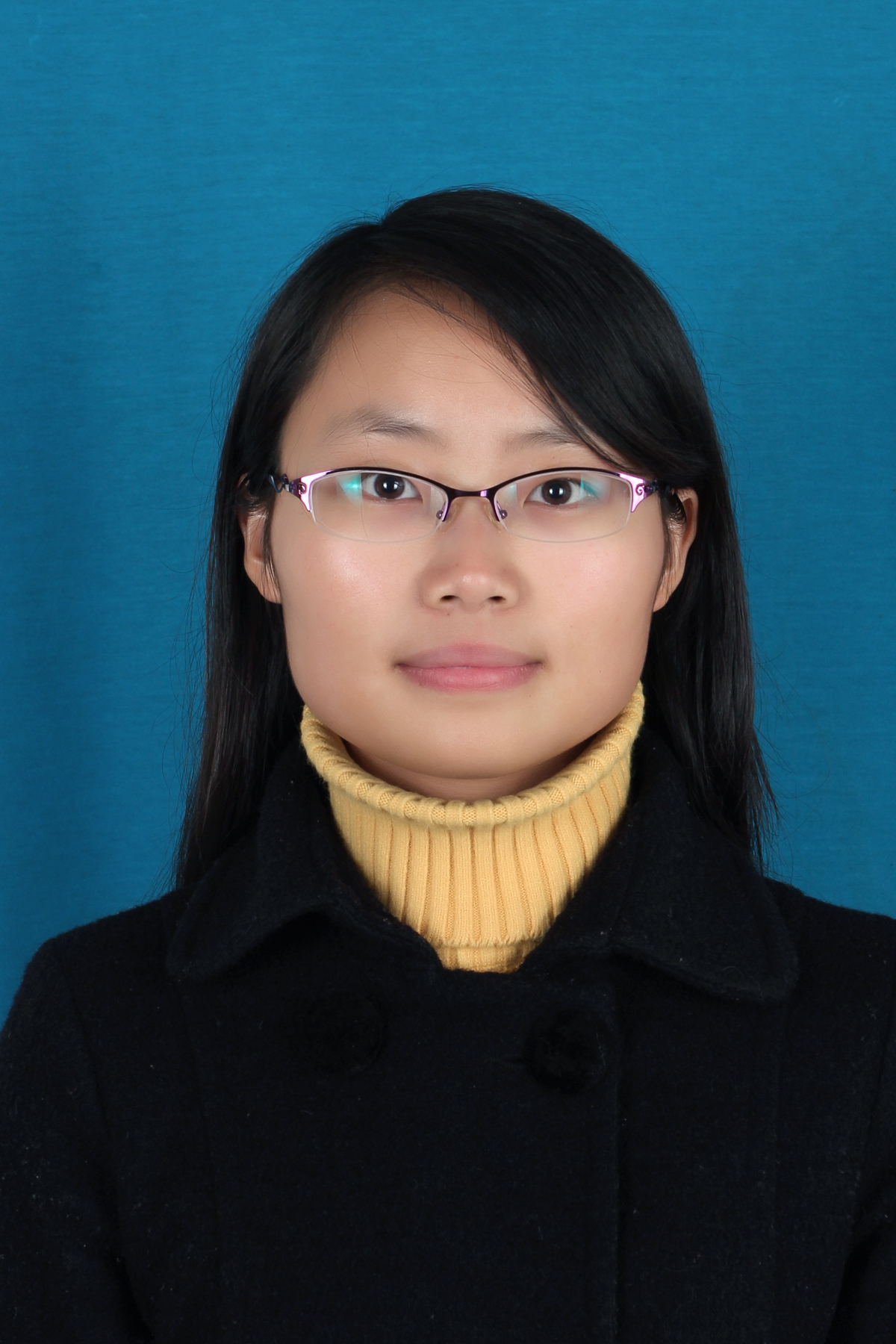}}]{QIN WANG}   (e-mail: 283252764@qq.com) is currently a lecturer at Guilin University of Electronic Technology, focusing on literature translation.
\end{IEEEbiography}
\vspace{-15 mm}
\begin{IEEEbiography}[{\includegraphics[width=1in,height=1.25in,clip,keepaspectratio]{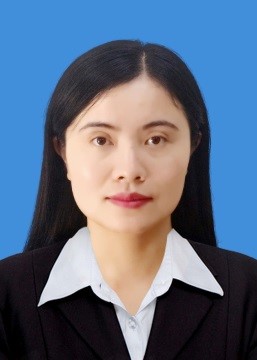}}]{QIN QIN}  (e-mail:qinqin@guet.edu.cn) is currently a senior engineer, and mainly studies image processing, big data, wireless sensors.
\end{IEEEbiography}
\vspace{-15 mm}
\begin{IEEEbiography}[{\includegraphics[width=1in,height=1.25in,clip,keepaspectratio]{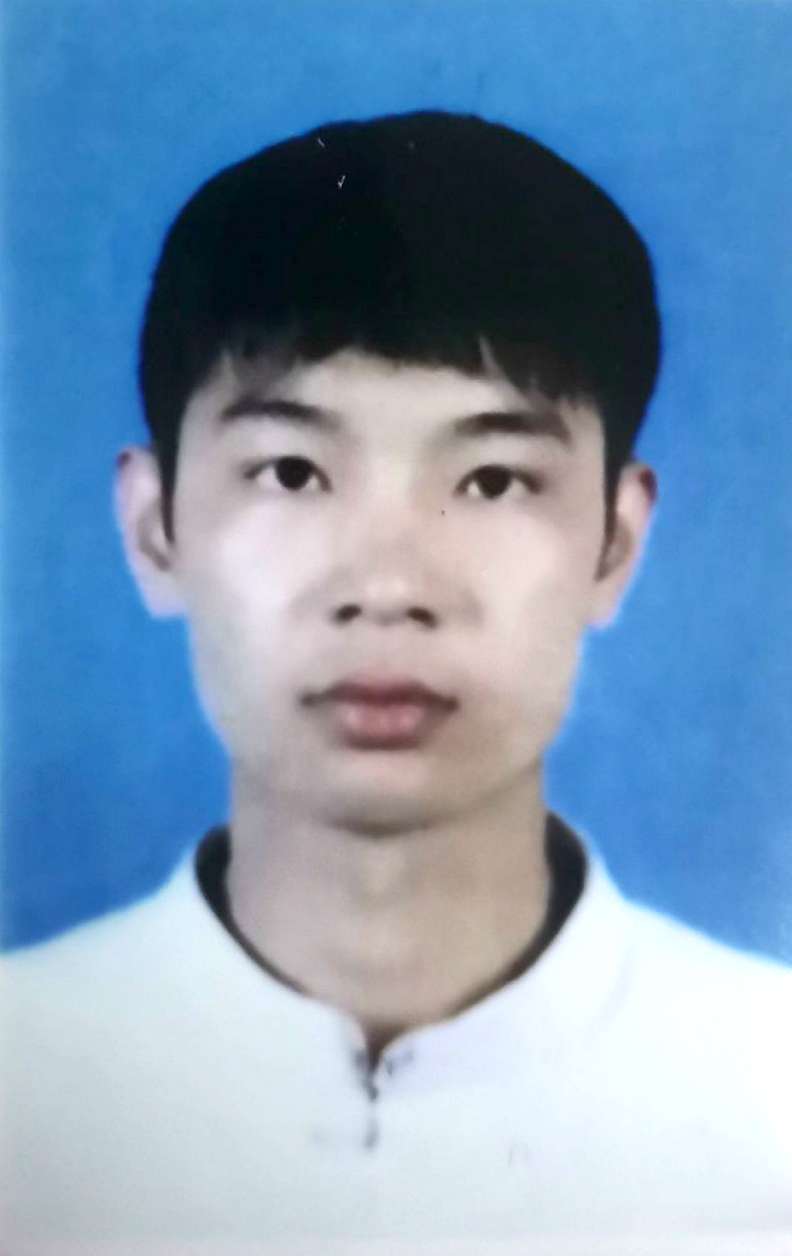}}]{HUANG-HE LI}  (e-mail:1556514980@qq.com) is currently a graduate student in the Ocean Engineering College of Guilin University of Electronic Technology. His research directions mainly include machine learning and microcontroller development.
\end{IEEEbiography}
\vspace{-10 mm}
\begin{IEEEbiography}[{\includegraphics[width=1in,height=1.25in,clip,keepaspectratio]{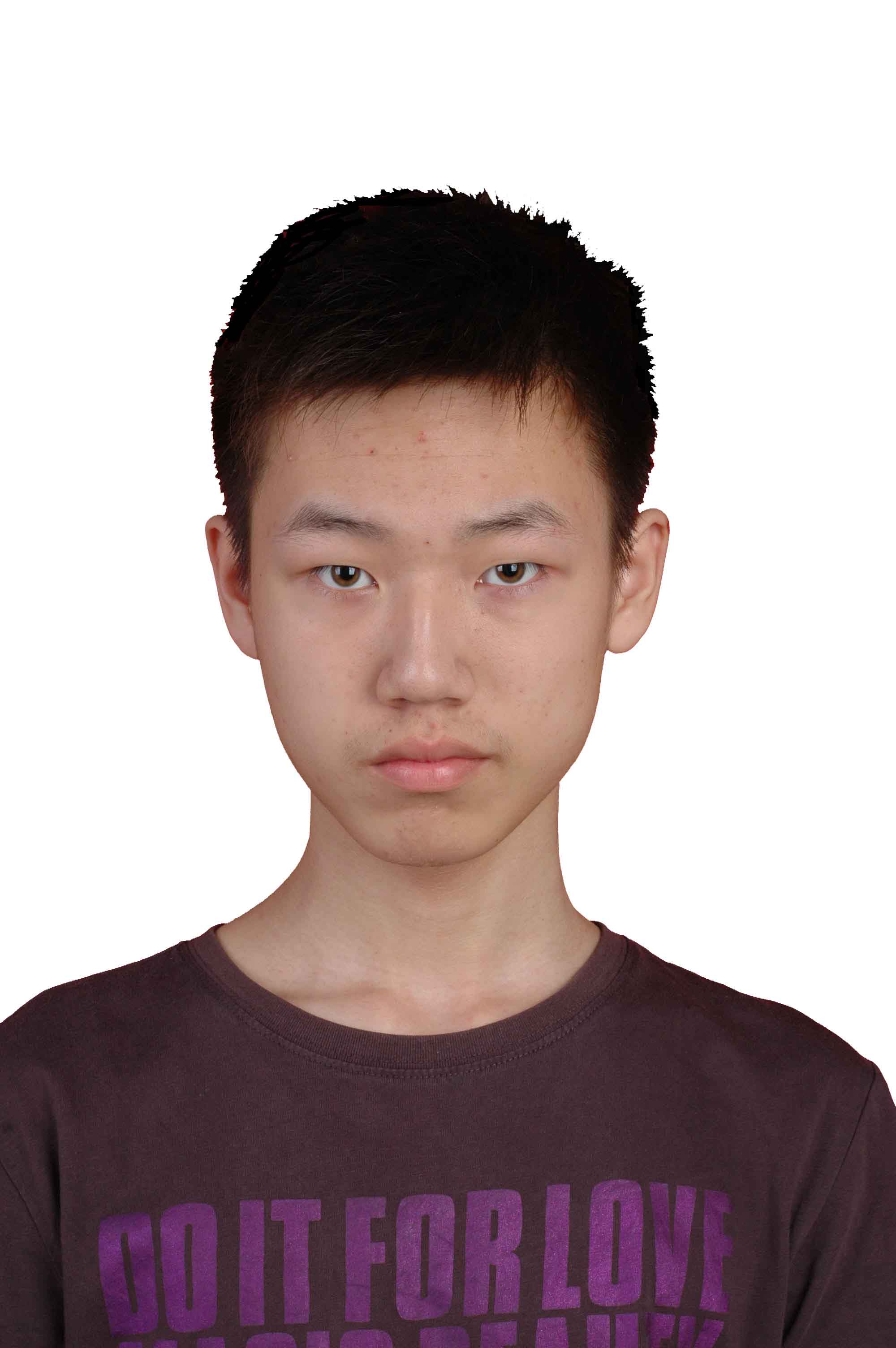}}]{PENG-FEI YE}  (e-mail:504551373@qq.com)  is currently a postgraduate student in the school of computer and information security of Guilin University of Electronic Science and technology. His research interests include image processing, image classification, and target detection.
\end{IEEEbiography}
\vspace{-10 mm}
\begin{IEEEbiography}[{\includegraphics[width=1in,height=1.25in,clip,keepaspectratio]{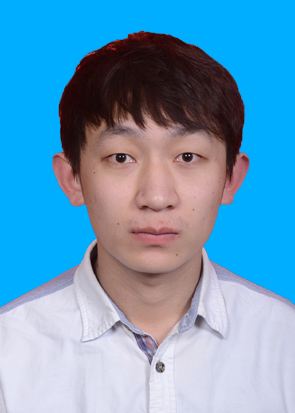}}]{YUE-PING HE}  (e-mail:553495998@qq.com)  is currently a postgraduate student in the school of Marine Engineering of Guilin University of Electronic Technology.His research interests include image processing, deep learning,instance segmentation.
\end{IEEEbiography}
\vspace{-20 mm}
\begin{IEEEbiography}[{\includegraphics[width=1in,height=1.25in,clip,keepaspectratio]{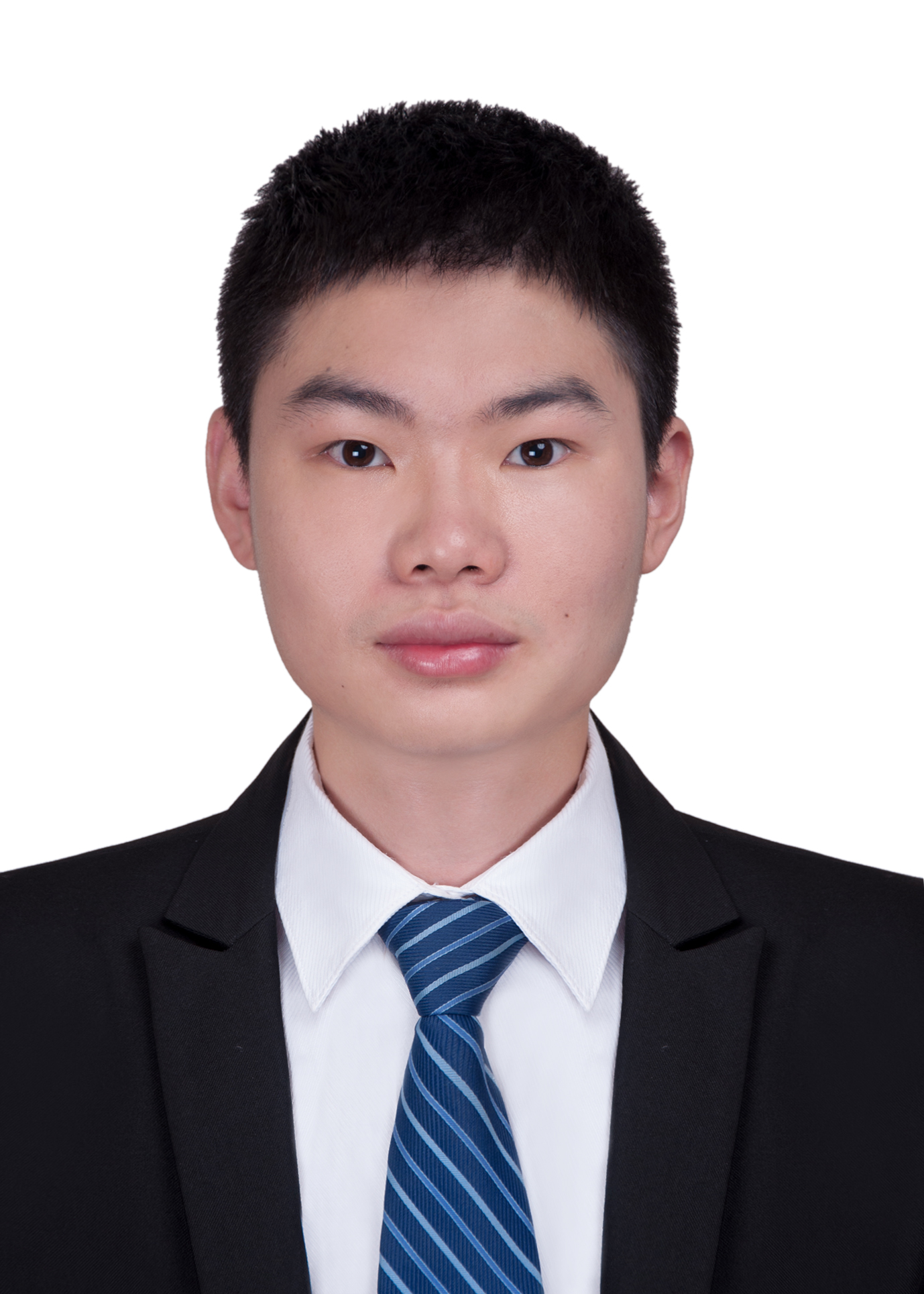}}]{JING ZENG}  (e-mail:962542336@qq.com)  is currently an intermediate engineer in the school of Marine Engineering of Guilin University of Electronic Technology.His research interests include 
Image processing, smart ships and ship building techniques
\end{IEEEbiography}

\newpage

\vfill

\end{document}